\documentclass[manuscript,screen,nonacm]{acmart}

\AtBeginDocument{%
  \providecommand\BibTeX{{%
    \normalfont B\kern-0.5em{\scshape i\kern-0.25em b}\kern-0.8em\TeX}}}

\usepackage{graphicx}
\usepackage{float}
\usepackage{caption}
\usepackage{todonotes}
\usepackage{subcaption}
\usepackage{comment}
\usepackage{hyperref}
\usepackage{algorithm2e}
\usepackage{xcolor}
\SetKwProg{func}{Function}{:}{end}
\DontPrintSemicolon

\usepackage{enumitem}
\setlist{leftmargin=10pt}

\begin{document}

\title{Cumulus: Blockchain-Enabled Privacy Preserving Data Audit in Cloud}
\author{Prabal Banerjee}
\email{mail.prabal@gmail.com}
\orcid{0000-0001-6142-9378}
\affiliation{%
  \institution{Avail and Indian Statistical Institute}
  \city{Kolkata}
  \state{West Bengal}
  \country{India}
}

\author{Nishant Nikam}
\affiliation{}
\email{nikam_nishant@outlook.com }

\author{Subhra Mazumdar}
\affiliation{%
  \institution{TU Wien and 
Christian Doppler Laboratory Blockchain
Technologies for the IoT}
  \city{Vienna}
  \country{Austria}
}
\email{subhra.mazumdar@tuwien.ac.at}
\orcid{0000-0002-3089-2535}

\author{Sushmita Ruj}
\affiliation{%
 \institution{University of New South Wales}
 \city{Sydney}
 \country{Australia}}
\email{sushmita.ruj@unsw.edu.au}
\orcid{0000-0002-8698-6709}

\renewcommand{\shortauthors}{Banerjee and Nikam, et al.}



\begin{abstract}
Data owners upload large files to cloud storage servers, but malicious servers may potentially tamper data. To check integrity of remote data, Proof of Retrievability (PoR) schemes were introduced. Existing PoR protocols assume that data owners and third-party auditors are honest and audit only the potentially malicious cloud server to check integrity of stored data. In this paper we consider a system where any party may attempt to cheat others and consider collusion cases. We design a protocol, Cumulus \footnote{Cloud storage has no relation to clouds in the sky! But since this type of cloud indicates fair weather, often popping up on bright sunny days, we chose this name to illustrate that our protocol would make the future of cloud storage brighter}, that is secure under such adversarial assumptions and use blockchain smart contracts to act as mediator in case of dispute and payment settlement. We use state channels to reduce blockchain interactions in order to build a practical audit solution. The security of the protocol has been proven in Universal Composability (UC) framework. Finally, we illustrate several applications of our basic protocol and evaluate practicality of our approach via a prototype implementation for fairly selling large files over the cryptocurrency Ethereum. We implement and evaluate a prototype using Ethereum as the blockchain platform and show that our scheme has comparable performance. 

\end{abstract}




\maketitle

%

\section{Introduction}

Recent
years have seen enormous amount of data generated and people using multiple devices connected to the Internet. To cater to the need for accessing data across devices, cloud storage providers have come up. Similarly, there has been an increase in Anything-as-a-Service(XaaS) which needs data to be uploaded and stored on remote servers. To ensure integrity of uploaded data, Proof-of-Storage algorithms have been proposed. With publicly verifiable Proof-of-Retrievability schemes, data owners can potentially outsource this auditing task to third-party auditors who send challenges to storage servers. The storage servers compute responses for each challenge and gives a response. By validating the challenge-response pair, the auditor ensures that the stored file is intact and retrievable. 

The inherent assumption in the existing PoR schemes is that the data owner is always honest. The server may be malicious and try to erase data in an attempt to reduce it's cost. The PoR schemes offer guarantees that if the server responds correctly to the challenge set, then the file is recoverable with overwhelming probability. The auditor is trusted with checking the integrity of the stored data without gaining access to the data itself, but previous studies showed that the auditor might gain access by carefully selecting the challenge set it sends to the server, thus acting maliciously. Also, data owner may refuse payment to server and auditor in a bid to cut cost and act maliciously.

Blockchain, a distributed tamper-resistant ledger, has seen several use cases lately. With smart contract support, arbitrary logic can be enforced in a distributed manner, even in the presence of some malicious players. These abilities are used to make blockchain a trusted party to resolve disputes. Several works have used blockchain as a judge to settle disputes among involved parties \cite{Bentov2014, bitcoin_incentivize, Dziembowski2018a}. Most major blockchain platforms have an inherent currency which is used to make meaningful monetary contracts among participants. 
Our objective in this paper is to address the problems discussed above, without having any trust assumptions. The proposed solution must be either efficient or comparable to the state-of-the-art.

\medskip
\subsection{Our Contribution}
In this paper we aim to propose a blockchain based proof-of-storage model and prove its security even if data owner is corrupted. We study the collusion cases and argue that our system is secure unless the data owner, auditor and cloud server together collude and act maliciously. We implement a prototype using a modified version of Ethereum. We use the inherent currency of the blockchain platform to model and enforce our incentive structure. To make minimum blockchain interactions, we use state channels and perform off-chain transactions. Our experiments show that our improved system has comparable performance. We summarize the contribution as follows:

\begin{itemize}
  \item We propose a blockchain based data audit model, \emph{Cumulus}, where the auditor can be a third party.
  \item We design state channel based audit protocol to minimize blockchain commits.
  \item We use blockchain-based payment to incentivize players in the system.
  \item We prove the security of our protocol in the \emph{UC} framework and show that the scheme is resilient to any form of collusion between participating entities. 
  \item We implement a prototype on modified Ethereum and show comparable performance with overhead of around 6 seconds per audit phase.
\end{itemize}

\section{Preliminaries and Background}

\subsection{Notations}
We take $\lambda$ to be the security parameter. An algorithm $\mathcal{A}(1^\lambda)$ is a probabilistic polynomial-time algorithm when its
running time is polynomial in $\lambda$ and its output $y$ is a random variable which depends on the internal coin tosses of $\mathcal{A}$. A function $f : \mathbb{N} \rightarrow \mathbb{R}$ is called negligible in $\lambda$ if for all positive integers $c$ and for all sufficiently large $\lambda$, we have $f(\lambda) < \frac{1}{\lambda^c}$. An element $a$ chosen uniformly at random from set $\mathcal{S}$ is denoted as $a \stackrel{R}{\leftarrow} \mathcal{S}$. We use a secure digital signature algorithm \texttt{(Gen,Sign,SigVerify)}, where \texttt{Gen()} is the key generation algorithm, \texttt{Sign()} is the signing algorithm and \texttt{SigVerify()} is the signature verification algorithm. We use a collision-resistant cryptographic hash function $H$.

\subsection{Bilinear Pairings}
\textbf{Definition:} Let $G_1, G_2$ be two additive cyclic groups of prime order $p$, and $G_T$ another cyclic group of order $p$ written multiplicatively. A pairing is a map $e : G_1 \times G_2 \rightarrow G_T$,  which satisfies the following properties:
\begin{itemize}
\item \textbf{Bilinearity:} $ \forall a,b \in \mathbb{Z}_p^*,\ \forall P\in G_1, \forall Q\in G_2:\ $ $$e\left(a P, b Q\right) = e\left(P, Q\right)^{ab}$$  where $\mathbb{Z}_p^* = \{ 1 \leq a \leq p-1 : gcd(a,p)=1\}$ with group operation of multiplication modulo $p$. 
\item \textbf{Non-Degeneracy:} $e(g_1, g_2) \neq 1_G$ where $g_1$ and $g_2$ are generators of $G_1$ and $G_2$ respectively and $1_G$ is the identity element in $G_T$. 
\item \textbf{Computability:} There exists an efficient algorithm to compute $e$.
\end{itemize}
A pairing is called symmetric if $G_1 = G_2$. When we use symmetric bilinear pairings, we refer to a map of the form $e : G \times G \rightarrow G_T$ with group $G$\'s support being $\mathbb{Z}_p$. \cite{Galbraith2008}


\subsection{Proofs of Retrievability}
PoR schemes are used to guarantee a client that her uploaded data stored with the server is not tampered. It was first introduced by Juels and Kaliski \cite{Juels2007}. In the setup phase, the client encodes file with erasure codes to get preprocessed file $F$, where each block of file $m_i$ is an element in $\mathbb{Z}_p$. It computes authenticator tags $\sigma_i$ for each block of $F$ and uploads $F$ to server along with the authenticators. During audit phase, the client sends random challenges to the server which acts as the prover and responds with a proof. The client verifies the proof and the server passes the audit if the verification goes through.

The correctness of the PoR algorithm ensures that an honest server always passes an audit, i.e., the challenge-response pair verification outputs $1$. The soundness property ensures that $F$ can be retrieved from a server which passes the audits with non-negligible probability.

There are mainly two types of PoR schemes: privately verifiable and publicly verifiable. In privately verifiable schemes, the client herself audits the server, or the auditor knows secret about the data. In publicly verifiable PoR schemes, any third party auditor can generate challenges and verify responses by knowing public parameters of the client. A PoR scheme is called privacy preserving if the responses to a challenge does not reveal any knowledge about the data.

\subsection{File Processing and Query Generation}
\textbf{File:} A file $F$ is broken into $n$ chunks, where each chunk is one element of $\mathbb{Z}_p$. Let the file be $b$ bits long. We refer to each file chunk as $m_i$ where $1 \leq i \leq n$ and $n = \lceil b/\lg p \rceil$. We use chunk and block interchangeably to refer to each file chunk.\\

\textbf{Query:} $\mathcal{Q} = \{(i,\nu_i)\}$ be an $l$-element set, where $l$ is a system parameter, $1 \leq i \leq n$ and $\nu_i \in \mathbb{Z}_p$. The verifier chooses an $l$-element subset $I$ of $[1,n]$, uniformly at random. For each $i \in I$, $\nu_i \stackrel{R}{\leftarrow} \mathbb{Z}_p$.

\subsection{Shacham Waters Public Verifiability Scheme} \label{shachamwaters}
We use the Shacham and Waters Compact Proofs of Retrievability scheme with public verifiability \cite{Shacham2008}, which uses symmetric bilinear pairings. Let a user have a key pair $K=(sk,pk)$ where $sk = x \in \mathbb{Z}_p$ and $pk = v = g^x \in G$. Let $u \in G$ be a generator. For file block $i$, authentication tag $\sigma_i = [H(i) u^{m_i}]^{sk}$. The prover receives the query  $\mathcal{Q} = \{(i,\nu_i)\}$ and sends back $\sigma \leftarrow \prod_{(i,\nu_i)\in \mathcal{Q}} \sigma_i^{\nu_i}$ and $\mu \leftarrow \sum_{(i,\nu_i)\in \mathcal{Q}} \nu_i . m_i$. The verification equation is : 
\begin{equation} \label{verEqSW}
   e(\sigma, g) \stackrel{?}{=} e(\prod_{(i,\nu_i)\in \mathcal{Q}} H(i)^{\nu_i} . u^{\mu}, v) 
\end{equation}

The scheme has public verifiability because to generate authentication tags the private key $sk$ is required. On the other hand, for the proof-of-retrievability protocol, public key $pk$ is sufficient. In this paper, we refer to this scheme as Shacham-Waters.

\subsection{Privacy Preserving Public Auditing for Secure Cloud Storage} \label{pppascs}
To achieve strong privacy guarantees, we use Privacy Preserving Public Auditing for Secure Cloud Storage scheme by Wang \textit{et al.} \cite{Wang2013}. Let $H_1:{\{0,1\}}^{*} \rightarrow G_1$ and $H_2:G_T \rightarrow \mathbb{Z}_p$ be hash functions and $g$ be a generator of $G_2$. Let a user have a key pair $K=(sk,pk)$ where $sk = x \in \mathbb{Z}_p$ and $pk = (v,g,u,e(u,v))$ such that $v=g^x$ and $u \stackrel{R}{\leftarrow} G_1$. For file block $i$ with identifier $id \stackrel{R}{\leftarrow} \mathbb{Z}_p$, authentication tag $\sigma_i = [H_1(W_i) u^{m_i}]^{sk}$, where $W_i=id||i$. The prover receives the query  $\mathcal{Q} = \{(i,\nu_i)\}$. It selects a random element $r \stackrel{R}{\leftarrow} \mathbb{Z}_p$ and calculates $R=e(u,v)^r$ along with $\gamma=H_2(R)$. Finally the prover sends back $(\mu, \sigma, R)$ where $\sigma \leftarrow \prod_{(i,\nu_i)\in \mathcal{Q}} \sigma_i^{\nu_i}$ and $\mu \leftarrow r + \gamma \times (\sum_{(i,\nu_i)\in \mathcal{Q}} \nu_i . m_i)$. The verification equation is : 
\begin{equation} \label{verEqPPSCS}
   R . e(\sigma^\gamma, g) \stackrel{?}{=} e((\prod_{(i,\nu_i)\in \mathcal{Q}} H_1(W_i)^{\nu_i})^\gamma . u^{\mu}, v) 
\end{equation}
The scheme is also publicly verifiable like Shacham-Waters protocol. Additionally, it is privacy preserving. In this paper, we refer to this scheme as PPSCS. 

\subsection{Blockchain}
Blockchain is a tamper-resistant, append-only distributed ledger. Apart from acting as a non-repudiable log, blockchain can host distributed applications and perform arbitrary functions in the form of smart contracts. First introduced in Bitcoin \cite{Satoshi2008}, it is a hash-linked chain of blocks, each block potentially containing multiple transactions. Participating nodes broadcast ledger updates in form of transactions. While there are various flavours of blockchain systems present based on the type of consensus protocol they use, we use a Proof-of-Work(PoW) based blockchain system. In a blockchain platform following PoW consensus protocol, special nodes, called miners, form blocks containing multiple transactions. The miners compete and solve a hash based challenge and the winner gets to propose the next block along with getting a mining reward. 

Ethereum \cite{Wood2014} is one of the most popular blockchain platforms supporting Turing-complete languages to write smart contracts. Ether is the cryptocurrency of the Ethereum platform and it is used to incentivize computations on the platform. The contracts are executed inside Ethereum Virtual Machine(EVM) which is uniform across all nodes, so as to have same output across the network. The amount of work done in terms of the number of operations done is calculated in terms of gas. A user submits transactions along with ethers to compensate for the work done by miners, according to the gas price. This acts as a transaction fee for the miners, and at the same time prevents running bad code like infinite loops which might harm the miners. Ethereum has a set of pre-compiled contracts which are codes running inside the host machine and not inside the EVM. Hence, the pre-compiled contracts cost less gas. Ethereum is open-source with an active community and has seen large scale adoption.

While active research is being performed to lower consensus time in public blockchain systems, traditional PoW chains need considerable time before a transaction reaches finality. Hence, it becomes hard to implement multi-commit protocols which are practical. For example, in Ethereum, the average block generation time is between 10-19 seconds. This delay might not be suitable for high frequency applications. On top of that, for each transaction to get mined, the user needs to incur additional cost in terms of transaction fee or gas costs. These problems make protocols like audit unsuitable as an auditor and server need to interact multiple times to exchange challenges and responses. A typical technique used to bypass these problems is performing off-chain transactions by opening state channels between pairs of users. The participants of a state channel exchange signed messages and perform on-chain transactions only when either they are finished with their interaction or some dispute arises. The blockchain either saves the final states of the participants or resolves disputes, whichever applicable. This reduces time and cost for the users. 



\section{Related Work}
\textbf{On Cloud Storage:} The idea of auditing cloud storage servers was first introduced by Ateniese \textit{et al.}\cite{Ateniese2009}, who defined Provable Data Possession (PDP) model. Juels and Kaliski \cite{Juels2007} first described a PoR scheme for static data using sentinels. A similar construction was provided by Naor \textit{et al.} \cite{Naor2009} using MAC-based authenticators. A study on various variants of PoR schemes with private verifiability is done by Dodis \textit{et al.}\cite{Dodis2009}. First fully dynamic provable data possession was given by Erway \textit{et al.}\cite{Erway2009}. A secure distributed cloud storage scheme called HAIL (high-availability and integrity layer) is proposed by Bowers \textit{et al.}\cite{Bowers2009} which attains POR guarantees. Shacham and Waters provided a PoR construction using BLS signatures, which had both public and private verifiability.  Wang \textit{et al.} argued that the auditor can retrieve information about files and hence proposed a privacy-preserving data audit scheme in \cite{Wang2013}. For dynamic data, an ORAM-based audit protocol was given in \cite{Cash2017}. More efficient protocols were given in \cite{Dautrich2014, Wang2014SecurelyOE, Sengupta:2016:PVS:2897845.2897915, Wang2017OnlineOfflinePD, senguptaIEEECloud, FU201897, LI2020102545, RABANINEJAD2020102454}. Multiple server-based PoR schemes were formalized in \cite{Paterson2016MultiproverPO}. Multi-user based data integrity check with revocable user access was proposed in \cite{THOKCHOM2020102427}. An identity-based auditing scheme for medical data was proposed in \cite{XU2020102453}. Nayak and Tripathy \cite{8327915} came up with a protocol that guarantees a secure and efficient privacy-preserving provable data possession scheme (SEPDP) for cloud storage and extended it to support multiple data owners, batch auditing, and dynamic data operations. Their scheme fails to guarantee soundness and does not achieve zero-knowledge privacy. Ni et al. \cite{9252871} addressed this issue by proposing an IDentity-based Privacy-Preserving Provable Data Possession scheme (ID-P3DP) based on the RSA assumption for secure cloud storage. It supports data privacy preservation against third-party auditors and multi-user aggregate verification. Yang et al. \cite{9733365} propose an ID-based PDP with compressed cloud storage where cloud storage auditing can be achieved by using only encrypted data blocks in a self-verified way and original data blocks can be reconstructed from the outsourced data.

In OPOR\cite{armknecht2014outsourced}, the authors define a formal framework where the auditing task is outsourced and provide a construction based on Shacham-Waters. Although it talks about handling all possible collusion cases, the scheme is shown to be secure for arbitrary collusion in terms of file recovery and not reliability, i.e., a misbehaving auditor may be able to prove innocence if it colludes with a malicious data owner. Also, in OPOR, the file is assumed to be encrypted and uploaded by the data owner to the server as the scheme does not ensure the privacy of the file. It also fails to provide a payment mechanism or arbitration. For multi-user owned data integrity check, Yuan \textit{et al.} \cite{6848154} proposed a scheme that handled collusion between parties, but was shown to be flawed in \cite{ZHANG201568}. In our paper, we provide an end-to-end system that handles data privacy, payment, and arbitration and is resilient to arbitrary collusion between parties. 

\noindent\textbf{Auditing using Blockchain:} Wang et al. \cite{wang2020blockchain} proposed the notion of Non-Interactive Public Provable Data Possession (NI-PPDP) and designed a blockchain-based fair payment smart contract for cloud storage based on this primitive. However, this work still suffers from some limitations that it is not easy to trace the malicious participant and works if the cloud service provider does not collude with the auditor. Yue et al. \cite{yue2020blockchain} proposed a framework for decentralized Edge-Cloud Storage. Efficient verification is ensured by sampling verification and formulated rational sampling strategies. However, this work does not have any incentive mechanisms, and the computational cost and the communication cost cannot be evaluated. A blockchain-based efficient public integrity auditing scheme to resist misbehaved third-party auditor is proposed in \cite{li2021blockchain}. Here the user is required to check the behaviors of auditors for a longer duration compared with that of the data integrity auditing performed by the auditor. The protocol fails to work in the presence of a fully malicious cloud server. Zhang et al. \cite{zhang2021blockchain} used the blockchain to record the interactions among users, service providers, and organizers in the data auditing process as evidence. Moreover, the smart contract was employed to detect service disputes. Xie et al. \cite{xie2022novel} use smart contracts to improve the reliability and stability of audit results. However, the authors assume that the data owner is a trusted entity. In \cite{li2022pptps}, the authors proposed a privacy-preserving auditable service with traceable timeliness for public cloud Storage. However, it is assumed that the data owner does not collude with the cloud service provider. Additionally, the scheme works if the auditor does not collude with the data owner and the cloud. 

\noindent\textbf{On Storage with Blockchain:} In recent times, various blockchain-based cloud servers have come up. IPFS\cite{Benet2014} introduced a blockchain-based naming and storage system. Several other systems \cite{Wilkinson2014, Ali2016,Vorick2014} use the concept of cloud storage in a decentralized fashion in a P2P network. In \cite{Ateniese2017}, the authors make the storage accountable and show how to integrate with Bitcoin. In \cite{Benet2018}, the designers use IPFS and cryptocurrency to make a storage-based marketplace. They use Proof of Replication to enforce storage among their peers. SpaceMint\cite{spacemint} introduced a new cryptocurrency that adapts proof of space, and also proposed a different blockchain format and
transaction types. Moran \textit{et al.}\cite{spacetime} introduced Proofs of Space-Time (PoSTs) and implemented a practical protocol for these proofs. An in-depth analysis of Proof of Replication mechanisms was done in \cite{Fisch2018PoRepsPO}. While the use of blockchain as an enforcer and incentive distribution mechanism was tapped in these works, most works did not consider fairness among the services offered by the parties. IntegrityChain, a decentralized storage framework supporting provable data possession (PDP) based on blockchain was proposed in \cite{9060815}. Fairness in trading is ensured as a party loses coins upon misbehaving and earns if it behaves honestly.

Several works use deduplication technology by performing auditing on one copy of multiple same data, thereby significantly reducing storage overhead. In \cite{xu2020blockchain}, a client-side data deduplication scheme based on bilinear-pair techniques has been proposed. Blockchain records the behaviors of entities in both data outsourcing and auditing processes, ensuring the credibility of audit results. The paper lacks a discussion on fairness. Tian et al. \cite{9542861} proposed a blockchain-based secure deduplication scheme in decentralized storage. The scheme is based on the double-server storage model to achieve efficient space-saving while protecting data users from losing data under a single point of failure and duplicate-faking attack. Transparent integrity auditing was introduced in \cite{9690028} based on the blockchain. The authors have constructed a secure transparent deduplication scheme that supports deduplication over encrypted data.  Li et al. \cite{li2022blockchain} proposed a secure transparent deduplication scheme based on the blockchain that supports deduplication over encrypted data and enables users to attest the deduplication pattern on the cloud server. Another paper \cite{10018273} discusses a scheme that bridges secure deduplication and integrity auditing in encrypted cloud storage. However, all these works consider the cloud server to be semi-trusted or the third-party auditor to be trusted. In \cite{mishra2022enabling}, a blockchain-based secure decentralized public auditing in decentralized cloud storage has been proposed. The authors have used redactability for blockchain to handle security issues. Additionally, the model uses an efficient deduplication scheme to attain adequate storage savings while preserving the users from data loss due to duplicate faking attacks. Liu et al. \cite{LIU2023103718} propose a blockchain-based compact audit-enabled deduplication scheme in decentralized storage. The protocol adopts an aggregatable vector commitment to generate audit tags to overcome the low coupling problem between deduplication and auditing. The drawback of the protocol is that it considers the storage service provider to be honest-but-curious and does not work if the party is malicious.

\begin{figure}[H]
  \centering
  \captionsetup{justification=centering,font=scriptsize}
  \includegraphics[width=0.8\textwidth]{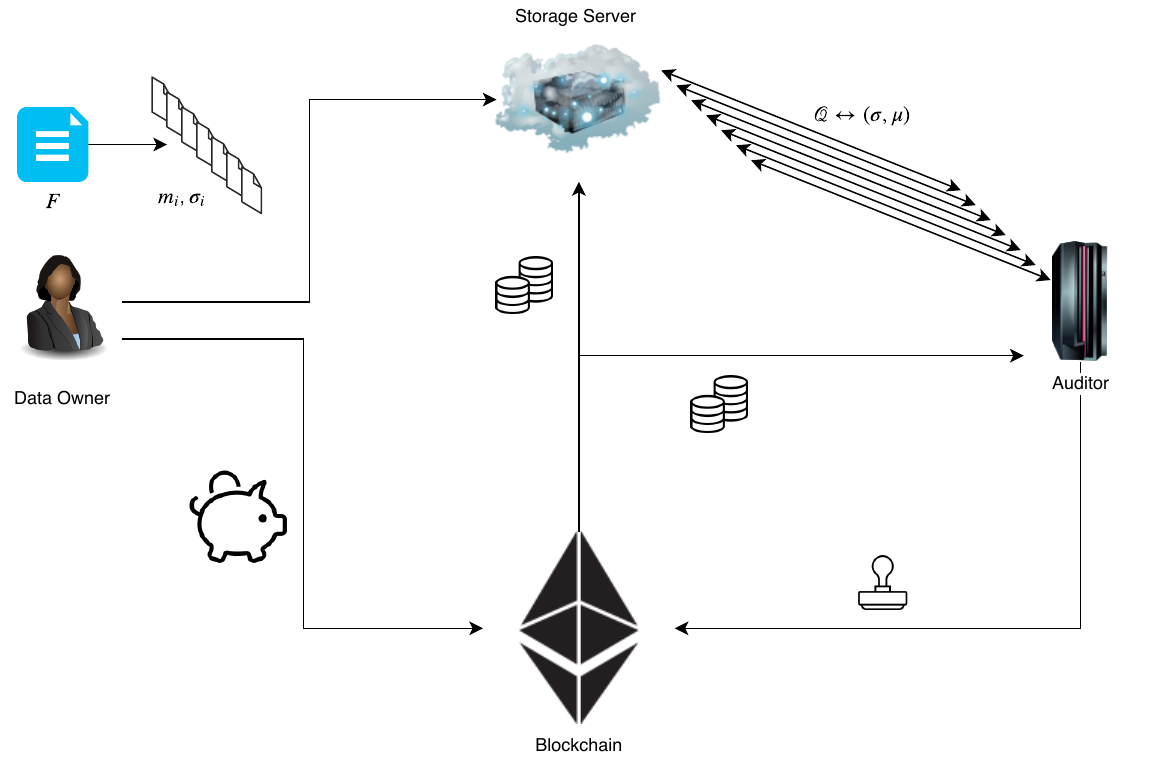}
  \caption{Overview of Interactions between Parties}
  \label{fig:overview}
\end{figure}

\section{Protocol Overview}
In this section we give an overview of the protocol and outline the deliverables we seek out of it. Figure \ref{fig:overview} provides an outline of our protocol.
\subsection{System Model}
The protocol has three entities: \emph{data owner, cloud storage server and auditor}. We use the blockchain layer as an arbitrator in case of disputes, using the Turing-complete capability of the blockchain platform to codify the actions in case of dispute. The native currency of the platform is used to distribute and control incentives. 
The immutability of the blockchain helps keep log of audit results and provides a transparent infrastructure without sacrificing on privacy. We assume that the \emph{data owner} sets up the smart contract and the \emph{cloud server} and \emph{auditor} interact only if they agree to the terms of the contract. If using a permissionless blockchain, any entity can run the blockchain network and the smart contract can be deployed on it. In case a private blockchain network is used, we think the storage providers, auditors and data owners will participate in the blockchain. The work done by the participants are incentivized in terms of gas cost and hence there might be entities who participate in the network just to gain that incentive.

\subsection{Adversarial Model \& Assumptions}

We would call a user honest, if she follows the protocol. Otherwise, we would call her malicious. A malicious user can deviate arbitrarily.

An adversary is a polynomial-time algorithm that can make any user malicious at any point of time, subject to some upper bound. Our adversary is dynamic in nature, that is, it can select its target based on current configuration of the system. It can make coordinated attacks, that is, it can control the malicious users and send/receive messages on their behalf. It can, of course, make a malicious user isolated and prescribe arbitrary instructions for her to perform. \cite{Algorand}

However, the adversary cannot break cryptographic primitives like hash functions or signatures, except with negligible probability. It cannot interfere with honest users or their exchanges. Apart from this, we make the following assumptions:\\
(i) Among the peers in the blockchain, the adversary can only corrupt upto the bound of the underlying consensus protocol. For example, in PoW based blockchains the bound is 49\%, and for PoS based blockchains the bound is 33\%.\\
(ii) The three parties apart from the blockchain - owner, server and auditor - cannot be corrupted together. At most two of the three parties can be malicious at any point of time. \\
(iii) We assume that the adversary will not corrupt without sufficient incentive. We think of the adversary as a rational adversary.
\smallskip
\subsection{Security Goals}
We define the security goals that must be realized by \emph{Cumulus}.
\begin{itemize}
  \item \textbf{Authenticity:} The authenticity of storage requires that the cloud server cannot forge a valid proof of storage corresponding to the challenge set $\mathcal{Q}$ without storing the challenged chunks and their respective authentication tags untampered, except with a probability negligible in $\lambda$.
  \item \textbf{Extractibility:} The extractibility property requires the \texttt{FetchFile()} function to be able to recover the original file when interacting with a prover that correctly computes responses for non-negligible fraction of the query space.
  \item \textbf{Privacy:} The privacy of audit requires the auditor not to learn any property of the stored file chunks $m_i$. The auditor generates queries to receive response. The auditor should not be able to derive $m_i$, for any $i$, from the response.
  \item \textbf{Fairness:} We notice that the cloud server and auditor offer services in exchange for payment from the data owner. The fairness property would require the following :
  \begin{itemize}
  \item If the cloud server stores $m_i$, $\forall i$, then it receives adequate incentive. If it fails to keep the files intact, it gets penalized. 
  \item If the auditor generates queries correctly, verifies responses and submits aggregated response to blockchain, then it receives an appropriate incentive. If not, it gets penalized. 
  \item If the data owner gets services from cloud server and auditor as intended, then it has to pay according to the agreement. If she incurs losses due to a malicious party, she will be paid for the damage.
  \end{itemize}
\end{itemize}
\begin{figure*}[ht]
\begin{subfigure}{0.5\textwidth}
    \centering
	\includegraphics[width=\textwidth]{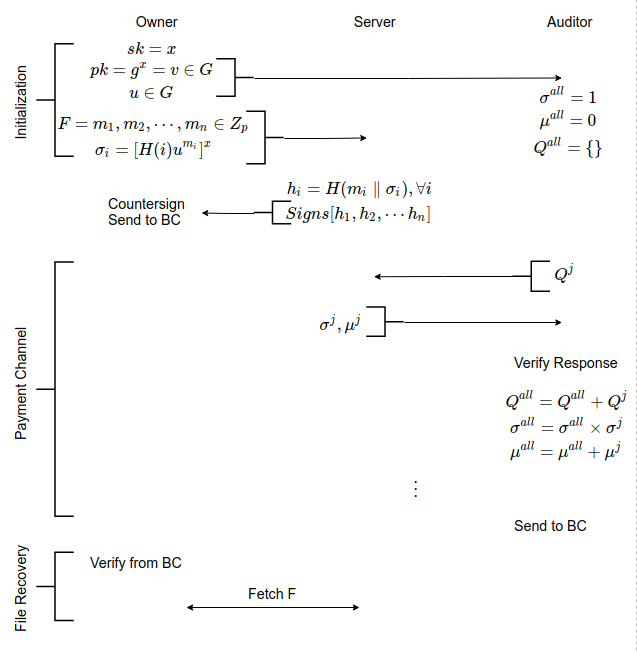}
    \caption{Flow of the Protocol in AuB}
    \label{fig:flow}
\end{subfigure}
\begin{subfigure}{0.5\textwidth}
    \centering
	\includegraphics[width=\textwidth]{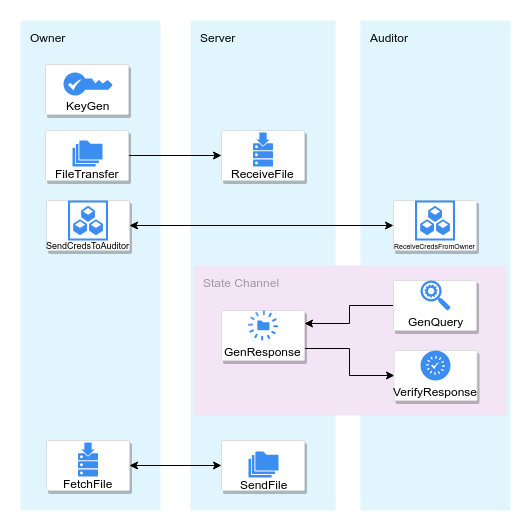}
    \caption{Generic Data Flow of the Protocol}
    \label{fig:generic}
\end{subfigure}
\caption{Construction Overview}
\end{figure*}
\subsection{Protocol Phases} \label{phases_subsection}
We break our protocol into four different phases. Let us outline the details of each of these phases. 
\medskip
\newline\textbf{Phase 0: Initialization Phase}
\begin{itemize}
    \item \textbf{KeyGen}: Initialized by the data owner, this algorithm generates a random public-private key-pair $K=(sk,pk)$ and public parameters 
    based on the security parameter $\lambda$.
    
    \item \textbf{RegisterOwner}: A new data owner uses this function to setup her account. She deposits prerequisite money to initialize her account with, which is used for future payments. The identity information, \texttt{ownerID}, will be used to authorize all further transactions by this data owner. She also submits her public key $pk_o$ which is used to verify signed messages submitted by the owner.
    
    \item \textbf{RegisterServer}: An existing data owner uses this function to supply identity information about the publicly known
    server, \texttt{serverID}, and public key $pk_s$ of the server with whom she wants to store her data. 
    
    \item \textbf{RegisterAuditor}: This function is called by the operative data owner to specify the publicly known auditor, \texttt{auditorID}, she wants to assign. If the owner wants the selected auditor to only audit a particular server, the owner may additionally supply that information. 
\end{itemize}
\smallskip
\textbf{Phase 1: Owner - Server}
\begin{itemize}
  
  \item \textbf{FileTransfer}: The owner divides the file F into $n$ blocks. Let $F = m_1, m_2, \cdots, m_n \in \mathbb{Z}_p$. She generates authentication tags, $\sigma_i$, $1 \leq i \leq n$ and calculates hash $h_i = H(m_i \parallel \sigma_i)$ for $1 \leq i \leq n$. She sends $(m_i,\sigma_i), \forall i$, to the server. She receives $([h_1', h_2', \cdots h_n'], t)$ from server, where $t$ is the signature tag. She checks \texttt{SigVerify}$([h_1', h_2',$ $\cdots, h_n'], t, pk_s)=1$ and $h_i=h_i' (\forall i)$. Then, she sends \texttt{Sign}$(([h_1',$ $h_2', \cdots, h_n'], t), sk_o)$ to the blockchain. 
\end{itemize}
\smallskip
\textbf{Phase 2: Server - Auditor}
\begin{itemize}
  \item \textbf{OpenChannel}: This function Collects deposit from server and auditor and opens up a state channel between them to interact off-chain. It freezes the owner money to pay necessary parties once channel is closed.
  
  \item \textbf{GenQuery}: This function generates an audit query for the auditor based on the randomness derived from the last block of the blockchain. The query is then sent to the server for response.
  
  \item \textbf{GenResponse}: Given a query, this function generates an audit response. The server sends the response to the auditor.
  
  \item \textbf{Verify}: Given a response, this function verifies whether the response is correct or not. Based on this outcome of verification, we proceed with the next set of challenges.
  
  \item \textbf{CloseChannel}: This function receives aggregated \\ challenge-response along with the complaint, if any. It verifies whether the queries were valid and responses pass the audit. In case of complaint, it punishes the guilty party, else, it pays server and auditor as per norms of payment.
\end{itemize}
\smallskip
\textbf{Phase 3: Owner - Server}
\begin{itemize}
  \item \textbf{FetchFile}: The data owner retrieves the stored file from the server using this function. 
\end{itemize}
 For a protocol to fit into our framework, the PoR scheme has to be publicly verifiable and needs to produce short aggregated proofs. Although multiple PoR schemes have our required properties, we chose Shacham-Waters in our first protocol named \textit{Audit using Blockchain(AuB)}. Shacham-Waters have complete security proofs along with practical overhead in terms of implementation. Also, it uses Homomorphic Linear Authenticators (HLA) which helps us have a very concise proof, which can be submitted to the ledger for verification upon closing state channel. The major drawback of Shacham-Waters is that it lacks privacy. Attacks have been shown that reveal parts of data from audit proofs. Hence, we further define a privacy preserving audit scheme using blockchain named \textit{Privacy Preserving Audit using Blockchain(PPAuB)} using PPSCS which gives privacy guarantees by random masking. We discuss the designs of \textit{AuB} and \textit{PPAuB} in this section. Figure \ref{fig:generic} provides a generic overview of the data flow in our construction. 

We assume that the server and auditor are known entities in the system, i.e., their public keys, identities and addresses are known throughout the system. Also, we assume the server and auditor has some coins deposited in the system which can be used to penalize them in case of misbehavior. For simplicity, we assume a single file $F$ uploaded by a single owner $\mathcal{O}$ to server $\mathcal{S}$. We assume $\mathcal{O}$ authorizes auditor $Au$ as the third-party auditor for performing audits. $\mathcal{O}$ might as well act as $Au$ herself and perform the audit protocol. Our security assumptions allow such a case because the protocol is resilient against collusion by owner and auditor.

\section{Our Security Model}
In this section, we model the security properties of \emph{Cumulus} in the global Universally Composable (UC) framework \cite{canetti2000security}  and discuss the security guarantee of our proposed protocol in this framework. We assume the set of parties involved in the protocol is fixed and the public keys of all parties are known. For our protocol, we consider static corruption where a PPT (probabilistic polynomial time) adversary $\mathcal{A}$ can corrupt any party at the beginning of the protocol. Once a party is corrupted, $\mathcal{A}$ can read the internal state, as well as all of the incoming and outgoing messages, of that party.  

We analyze the security of \emph{Cumulus} in the real world and simulates its execution in an ideal world. We define an ideal functionality that acts like a trusted party and all the parties interact with the ideal functionality. The ideal functionality is like an abstraction of our proposed protocol where we prove the security properties realized by our protocol. We also define an ideal world adversary, $Sim$, that attacks the ideal world functionality in the same way like the adversary $\mathcal{A}$ attacks the protocol execution in the real world. We also define a PPT environment $\mathcal{Z}$ that sends and receives information in both the real and ideal world. Our protocol is UC-secure if the environment $\mathcal{Z}$ can distinguish its interaction with the real world and ideal world with negligible probability. We discuss next the communication model, the global ledger functionality $\mathcal{L}$, and the global random oracle before describing the
ideal functionality  $\mathcal{F}^{\mathcal{L}}_{\textsf{Audit}}$ for file integrity audit.

\subsection*{Communication Model}
It is assumed that the communication between parties happen in a synchronized fashion, with protocol execution taking place in rounds. All honest parties are assumed to follow an ideal global clock $\mathcal{G}_{clock}$ \cite{badertscher2017bitcoin} which keep tracks of the time of each round. Offline communication between honest parties is assumed to occur via ideal functionality $\mathcal{F}_{smt}$, which ensures secure message transmission. Message send by party $P$ at round $t$ reaches party $Q$ at $t+1$. An adversary gets to know when the message is being sent out but doesn't get to know the content of the message. 
Messages exchanged between parties and environment $\mathcal{Z}$ or parties and the adversary is assumed to take 0 rounds for transmission.

\subsection*{Global random oracles}

 We will use a global random oracle $\mathcal{H}$  \cite{Dziembowski2018a} which is accessible by several instances of the protocol. It is like a log file, where given an input, an output is randomly generated. If the oracle is queried more than once on the same input, it returns the same output without regenerating one. 

\subsection*{Global Ledger Functionality}
We define a ledger functionality $\mathcal{L}$ that captures the basic functionality of transferring coins between parties and also locking and unlocking coins from a smart contract \cite{dziembowski2019perun}. We assume that $\mathcal{L}$ has access to the global random oracle $\mathcal{H}$. The ledger functionality is used both in the real and ideal world and can be accessed by multiple instances of the protocol, hence it is called the global ideal functionality \cite{Dziembowski2018a}. 

We define $\mathcal{L}$ in Fig.\ref{ledger:fig} as stated in \cite{Dziembowski2018a}. A list $L$ is maintained that keeps track of the coins locked in the contract denoted by contract identifier $id$. The \emph{update} function maintains the balance of each party $P_i$ in the local list $L$. Any party $P_i$ present in the instantiation of the contract $id$, locks coins in the contract by calling module \emph{freeze}. If the balance of party $P_i$ in $L$ is more than 0, then the coins can be frozen in the contract. Else, an error \emph{no balance} is returned. Upon calling the module \emph{unfreeze}, the parties can unlock their locked coins from contract $id$. Upon querying the latest state of $\mathcal{L}$, the global random oracle $\mathcal{H}$ is called with $L$ as the input. Basically, this value acts as the $seed$ which is used by the auditor in generating queries for the contract $id$.

 \begin{figure*}[!ht]
	\begin{center}
		
		\fbox{

			\begin{minipage}{0.9\textwidth}
			
				Functionality $\mathcal{L}$, running with a set of parties $\mathcal{P}_1,\dots, \mathcal{P}_n$ stores the 
				balance $p_i \in \mathbb{N}$  for every party $p_i, i \in [n]$ and a partial function $L$ for frozen cash. It accepts queries of the following types:
			\begin{itemize}
			    \item \textbf{Update Funds} Upon receiving message $(update, \mathcal{P}_i, p)$ with $p \geq 0$ from $\mathcal{Z}$ 
				set $p_i = p$ and send $(updated, \mathcal{P}_i, p)$ to every entity.

				\item \textbf{Freeze Funds} Upon receiving message $(freeze, \mathcal{P}_i, p)$ from an ideal functionality of 
				session $id$ check if $p_i > p$. If this is not the case, reply with $(nofunds, \mathcal{P}_i, p)$. 
				Otherwise set $p_i = p_i - p$, store $(id, p)$ in $L$ and send $(frozen, id, \mathcal{P}_i, p)$ to every entity.

				\item \textbf{Unfreeze Funds} Upon receiving message $(unfreeze, id, \mathcal{P}_j, p)$  from an ideal functionality
				of session $id$, check if $(id p') \in L$ with $p' \geq p$. 
				If this check holds update $(id, p')$ to $(id; p' - p)$, set $p_j = p_j + p$ and send 
				$(unfrozen, id, \mathcal{P}_j, p)$ to every entity.
				\item \textbf{Latest State} Upon receiving message $(get\_latest\_state,id)$ from an ideal functionality
				of session $id$, call the global random oracle $\mathcal{H}$ on input $L$, and return $\mathcal{H}(L)$.  
			\end{itemize}

		\end{minipage}
		}	
	\end{center}
	\caption{Global ledger functionality $\mathcal{L}$}
 \label{ledger:fig}
\end{figure*}

\subsection*{Ideal functionality for file integrity audit}
The ideal functionality $\mathcal{F}^{\mathcal{L}}_{\textsf{Audit}}$ shows how a file owner $\mathcal{O}$ shares a file with a cloud server $\mathcal{S}$, and later entrusts an auditor $Au$ with the task of guaranteeing that the server has stored the file without deleting or modifying it and $\mathcal{O}$ can retrieve it later as well. 

$\mathcal{F}^{\mathcal{L}}_{\textsf{Audit}}$ has four phases as defined in Fig. \ref{fig:ideal}: in the \emph{initialize} phase, the file is transferred from $\mathcal{O}$ to server $\mathcal{S}$ and, $\mathcal{O}$ locks coins $p_{\mathcal{S}}+p_{Au}$ in the ledger $\mathcal{L}$. $p_{\mathcal{S}}$ and $p_{Au}$ are the coins to be paid to $\mathcal{S}$ and $Au$ upon successful completion of service. In the \emph{channel open phase}, the auditor initiates the opening of a payment channel by sending a request to the server. The purpose of the payment channel is to ensure off-chain auditing of files and make a promise of payment if the audit is successful. If the server does not agree to open the channel, then coins locked by $\mathcal{O}$ are refunded. If both $Au$ and $\mathcal{S}$ are willing to open the channel, then they individually lock coins $a$ and $b$ respectively in the channel where $a<p_{Au}$ and $b<p_{\mathcal{S}}$. If $Au$ (or $\mathcal{S}$), misbehaves then it will lose $a$ (or $b$) coins and the coins will be used for compensating $mathcal{O}$ and $\mathcal{S}$ (or $Au$). 

In the \emph{query and response phase}, auditor sends a query $Q$. $\mathcal{F}^{\mathcal{L}}_{\textsf{Audit}}$ checks if $Q$ is generated from $seed$. It retrieves the seed by querying $\mathcal{L}$ for latest state. If the $Q$ is an invalid one, $Au$ is penalized. Else, $\mathcal{F}^{\mathcal{L}}_{\textsf{Audit}}$ waits for a response from $\mathcal{S}$. if the latter sends abort, then it means that $\mathcal{S}$ cannot answer the query and hence it loses $b$ coins. If the response is correct, then nothing is done. If the response is wrong, then $\mathcal{S}$ is penalized. If after receiving the response, $Au$ sends abort, then $a$ coins are used for compensating the server and owner. In \emph{channel close} phase, if $Au$ sends abort, it means that it had colluded with $\mathcal{S}$ and trying to send a wrong query-response set and hence, $Au$ is penalized. If this is not the case, then $\mathcal{F}^{\mathcal{L}}_{\textsf{Audit}}$ checks if the number of queries made is at least the $query\_count$. If this holds, $Au$ receives $p_{Au}$ coins and $\mathcal{S}$ receives $p_{\mathcal{S}}$ coins, marking the successful completion of file auditing. If the number of queries generated is less than the threshold, then $Au$ is penalized.

\begin{figure*}[!ht]
	\begin{center}
		\fbox{
			\begin{minipage}{1\textwidth}\
			
				The ideal functionality $\mathcal{F}^{\mathcal{L}}_{\textsf{Audit}}$ (in session $id$) interacts with an owner
				$\mathcal{O}$, a server $\mathcal{S}$, an auditor $Au$, the ideal
				adversary $Sim$ and the global ledger $\mathcal{L}$. Variables used:  $channel=\phi$.\\
				\textbf{(Initialize)}	
				\begin{itemize}
					\item Upon receiving $(store, id, F, p_S, p_{Au}, query\_count, n,  id_{Au}, id_{\mathcal{S}})$ from $\mathcal{Z}$,  leak $(store,id,|F|, query\_count,n, p_S, p_{Au},  \\ id_{\mathcal{O}}, id_{\mathcal{S}}, id_{Au})$ to $Sim$. Store $F$ as file chunks $m_i, \forall i \in [1,n]$ and store $query\_count$. 
					\item Send $(send,id, F,n,id_{\mathcal{S}})$ and $(auth,id)$ to $\mathcal{Z}$. If $(abort,id)$ is received from $id_{\mathcal{S}^*}$, then terminate the protocol.
				   
				\item Call $\mathcal{L}$ with $(freeze, id,p_S+p_{Au}, id_{\mathcal{O}})$.  $\mathcal{L}$ returns $(frozen, id, p_S+p_{Au},id_{\mathcal{O}})$ to $\mathcal{F}^{\mathcal{L}}_{\textsf{Audit}}$.
    \end{itemize} 
				\textbf{(Server-Auditor: Opening Channel)}
				\begin{itemize}

					\item Upon receiving $(open\_channel, id, a, id_{\mathcal{S}})$, forward the instruction  to $\mathcal{Z}$. If $(open\_channel, id, b, id_{Au})$ is received, leak the message of channel opening to $Sim$, call $\mathcal{L}$ with $(freeze, id,a, id_{Au})$ and $(freeze,id, b, id_{\mathcal{S}})$. $\mathcal{L}$ responds with $(frozen, id,a,id_{Au})$ and $(frozen,id, b,id_{\mathcal{S}})$. Set $channel=opened$, and leak $query\_count$ to $Sim$. If $\mathcal{S}^*$ sends $(abort,id)$, call $\mathcal{L}$ with $(unfreeze, id,p_S+p_{Au}, id_{\mathcal{O}})$ and abort. 
			       \item Send $(get\_latest\_state,id)$ to $\mathcal{L}$, gets the value returned by $\mathcal{L}$, and stores the value in variable $seed$.		
				\end{itemize}
            \textbf{(Server-Auditor: Query and Response)}
            
            Upon receiving $(query,id, Q,id_{Au})$:
            \begin{itemize}
            \item Leak $Q$ to $Sim$, forward the message to $id_{\mathcal{S}}$. Check if $Q$ can be generated from $seed$. If not, then call $\mathcal{L}$ with $(unfreeze, id,b+\frac{a}{2}, id_\mathcal{S})$ and $(unfreeze,id, p_S+p_{Au}+\frac{a}{2}, id_\mathcal{O})$. Send $(penalized,id,id_{Au})$, set $channel=closed$ and abort.
                
                \item Upon receiving $(response,id,\bar{m},id_{\mathcal{S}})$, construct set $\bar{m'}=\{m_i\}_{i \in Q \wedge m_i \in F}$, and check $\bar{m'}\stackrel{?}{=}\bar{m}$. 
            
                \begin{itemize}
                \item If this holds, then $Q_{all}= Q_{all}\cup Q$.
                 \item  If $\bar{m'}\neq m$, then call $\mathcal{L}$ with $(unfreeze,id, \frac{b}{2}+a, id_{Au})$ and $(unfreeze,id, \frac{b}{2}+p_S+p_{Au}, id_\mathcal{O})$. Send $(penalized,id,id_{\mathcal{S}})$ to $\mathcal{Z}$, set $channel=closed$ and terminate the protocol. 
                 \item If $(abort,id)$ is received from $id_{Au^*}$, then call $\mathcal{L}$ with $(unfreeze,id, \frac{a}{2}+b, id_{S})$ and $(unfreeze,id, \frac{a}{2}+p_S+p_{Au}, id_\mathcal{O})$. Send $(penalized,id,id_{Au})$ to $\mathcal{Z}$, set $channel=closed$ and terminate the protocol.
                    				
                 \end{itemize}

                     \item If $(abort,id)$ is received from $id_{\mathcal{S}^*}$, then call $\mathcal{L}$ with $(unfreeze,id, \frac{b}{2}+a, id_{Au})$ and $(unfreeze,id, \frac{b}{2}+p_S+p_{Au}, id_\mathcal{O})$. Send $(penalized,id,id_{\mathcal{S}})$ to $\mathcal{Z}$, set $channel=closed$ and terminate the protocol.

                \end{itemize}

				\textbf{(Server-Auditor: Close Channel)}
				
				\begin{itemize}
    \item Upon receiving $(abort,id)$ from $id_{Au^*}$:
    
\begin{itemize}
    \item call $\mathcal{L}$ with $(unfreeze, id,b+\frac{a}{2}, id_\mathcal{S})$ and $(unfreeze,id, p_S+p_{Au}+\frac{a}{2}, id_\mathcal{O})$. Send $(penalized,id,id_{Au})$ to $\mathcal{Z}$. Set $channel=closed$ and terminate.
\end{itemize}
					\item Upon receiving $(close,id,id_{Au})$: 
				\begin{itemize}
				\item  Leak the instruction to $Sim$. Check $\frac{|Q_{all}|}{query\_count}\stackrel{?}{\geq} 1$. If the inequality holds true, call $\mathcal{L}$ with $(unfreeze,id, a+p_{Au}, id_{Au})$ and $(unfreeze, id,b+p_S, id_\mathcal{S})$, and send $(check\_pass,id)$ to $\mathcal{Z}$.  $\mathcal{F}^{\mathcal{L}}_{\textsf{Audit}}$ sends ($send\_file,F,id_\mathcal{O})$, sets $channel=closed$ and terminate.
				   \item If the above step fails, call $\mathcal{L}$ with $(unfreeze, id,b+\frac{a}{2}, id_\mathcal{S})$ and $(unfreeze,id, p_S+p_{Au}+\frac{a}{2}, id_\mathcal{O})$. Send $(penalized,id,id_{Au})$ to $\mathcal{Z}$. Set $channel=closed$ and terminate.
					
				\end{itemize}

					\end{itemize}
						
			\end{minipage}
		}
	\end{center}
	\caption{Ideal Functionality $\mathcal{F}^{\mathcal{L}}_{\textsf{Audit}}$ for file integrity audit} 
 \label{fig:ideal}
\end{figure*}

\subsection*{Privacy and Security Properties}
We discuss how the ideal functionality
$\mathcal{F}^{\mathcal{L}}_{\textsf{Audit}}$ guarantees the following privacy and security properties:
\begin{itemize}
  \item \textbf{Authenticity:} $\mathcal{F}^{\mathcal{L}}_{\textsf{Audit}}$ checks the response $R$ corresponding to query $Q$ and matches each response with the file chunks it had received from $\mathcal{O}$. If the response is wrong, $\mathcal{S}$ is penalized.  

  \item \textbf{Extractibility:} Since a threshold $query\_count$ is set which is a significant fraction of total file size, if the auditing is successful, then $\mathcal{O}$ is assured that with high confidence, it will receive the correct file from $\mathcal{S}$. If the number of queries is less than the threshold, $Au$ is penalized by the ideal functionality. 
  
  \item \textbf{Privacy:} $Au$ does not come to know about the file content, it only sees whether the auditing was correct or not based on unfreezing of coins from $\mathcal{L}$.
  
  \item \textbf{Fairness:} If $Au$ and $\mathcal{S}$ fails to open channel, coins are refunded to $\mathcal{O}$. If auditor is malicious then $\mathcal{F}^{\mathcal{L}}_{\textsf{Audit}}$ penalizes the former and compensates $\mathcal{O}$ and $\mathcal{S}$ using $a$ coins. If server is malicious, then $\mathcal{F}^{\mathcal{L}}_{\textsf{Audit}}$ penalizes $\mathcal{S}$. If both $Au$ and $\mathcal{S}$ behave as per the protocol specification, then both are rewarded.
\end{itemize}

\section{Formal Description of the Protocol}

The protocol $\Pi_{Cumulus}$ defined in the $(\mathcal{F}_{ecdsa},\mathcal{G}^{\mathcal{L}}_{\textsf{JC}}, \mathcal{H})$-hybrid world. We define judge contract functionality  $\mathcal{G}^{\mathcal{L}}_{\textsf{JC}}$ before explaining the steps of the protocol. The ideal functionality $\mathcal{F}_{ecdsa}$ \cite{doerner2018secure} is used for providing the interface of signing and verifying messages sends by the parties. Since ECDSA signatures are strongly existentially unforgeable \cite{ecdsa}, we leverage on this property to argue for the security of the \emph{Cumulus}.


\begin{figure*}[!ht]
\label{formal}
\begin{center}
\fbox{
\begin{minipage}{1.04\textwidth}\
Consider an elliptic curve group $G$ of order $p$  with generator $g$, then:\\
\textbf{Keygen}: On receiving Keygen($id,G, g, q$) from party $P_{id}$:
\begin{itemize}
    \item Generate key pair $(sk,pk)$ where $sk=x: x\leftarrow \mathbb{Z}_p, pk=g^x$
    \item Return $(sk,pk)$ to $P_{id}$
\end{itemize}
\textbf{Sign}: On receiving $Sign(id,m)$:
\begin{itemize}
    \item Retrieve $sk_{id}$ corresponding to party $P_{id}$
    \item Choose $k\leftarrow \mathbb{Z}_q$, and compute $R=(r_x,r_y)=g^k$
    \item Compute $r= r_x \mod p$ and $s=k^{-1}(\mathcal{H}(m)+r.sk_{id}) \mod p$, by accessing global random oracle $\mathcal{H}$.
    \item Send $(r,s)$ to party $P_{id}$
\end{itemize}
\textbf{SigVerify}: On receiving $SigVerify(m,\sigma,pk)$:
\begin{itemize}
    \item Parse $\sigma$ and retrive $r$ and $s$

    \item Compute $K=g^{s(\mathcal{H}(m)+r}.pk$ 
    \item Given $K=(r_x',r_y')$, check if $r\stackrel{?}{=}r_x' \mod p$
\end{itemize}
\end{minipage}
}
\end{center}
\caption{Ideal Functionality $\mathcal{F}_{ecdsa}$ for ECDSA signature} 
\end{figure*}


\begin{figure*}[!ht]
\label{formal}
\begin{center}
\fbox{
\begin{minipage}{1.04\textwidth}\

The ideal functionality  $\mathcal{G}^{\mathcal{L}}_{\textsf{JC}}$ acts as a judge smart contract for session id $id$ and interacts
with the global $\mathcal{L}$ functionality and the parties $\mathcal{O}$, $\mathcal{S}$ and $Au$. \\

    \func{\texttt{RegisterOwner($pk_o, contractTerms$)}} {
        The Owner Deposits required money according to $contractTerms$.\;
        Return \texttt{OwnerID} to the Owner.
    }
    \func{\texttt{RegisterServer(OwnerID, ServerID)}} {
      Store the mapping \texttt{(OwnerID, ServerID)}\;
    }
    \func{\texttt{RegisterAuditor(OwnerID, AuditorID)}} {
      Store the mapping \texttt{(OwnerID, AuditorID)}\;
    }
  
    \func{\texttt{ReceiveSignedDigest($(m,t)$,$s$)}} {
        Require \texttt{SigVerify($m$,$t$,$pk_s$)=1} \;
        Require \texttt{SigVerify($(m,t)$,$s,pk_o$)=1} \;
    }

\textbf{(Initialize)}
\begin{itemize}
	\item Upon receiving $(init,id,pk_o, pk_{a},pk_s,u,v,g,contract \ terms,query\_count, id_{\mathcal{S}}, id_{Au})$, s tore the public keys $pk_o,pk_s$ and $pk_a$, parameters $v,u,g$ and $query\_count$. 
	\item Call \texttt{RegisterOwner} with input $pk_o$ and $contract \ terms$. Get $id_{\mathcal{O}}$. 
	\item Call \texttt{RegisterServer}, with input $id_{\mathcal{O}},id_{\mathcal{S}}$ and \texttt{RegisterAuditor}, with input $id_{\mathcal{O}},id_{Au}$.
	\item Output $(initialized, registered \ id_{\mathcal{O}}, registered \ id_{\mathcal{S}}, registered \ id_{Au})$.
	\end{itemize}
	\textbf{(File transfer from Owner to Server)}
\begin{itemize}

	\item Upon receiving $(verify,id,[h_1',h_2',\ldots,h_n'],t,\sigma_{file})$ from $\mathcal{O}$, call \texttt{ReceiveSignedDigest} with input $([h_1',h_2',\ldots,h_n'],t,\sigma_{file})$. If the verification passes, return $verified$.
\end{itemize}    
\textbf{(Server-Auditor: Opening Channel)}
\begin{itemize}
				    \item Upon receiving $(open\_channel,id, a,id_{Au},b,id_{\mathcal{S}})$ from $Au$, call $\mathcal{L}$ with $(freeze, id,a, id_{Au})$ and $(freeze,id, b, id_{\mathcal{S}})$. $\mathcal{L}$ responds with $(frozen, id,a,id_{Au})$ and $(frozen,id, b,id_{\mathcal{S}})$.
				    \item Upon receiving acknowledgement from $\mathcal{L}$, return $(opened\_channel,id)$.
				    
				\end{itemize}

\textbf{(Server-Auditor: Closing Channel)}
 
				\begin{itemize}
					\item Receive $(close,id, Q,sig_Q,(\mu,\sigma))$ from $Au$
					\begin{itemize}
						\item Parse $Q$ to get $(i,\nu_i)$, where $i$ is the index of the chunk. Check $e(\sigma,g)\stackrel{?}{=}e(\Pi_{(i,\nu_i)\in Q} H(i)^{\nu_i}u^{\mu},v)$. If this holds true and $\frac{|Q|}{query\_count}\geq 1$, call $\mathcal{L}$ with $(unfreeze,id, a+p_{Au}, id_{Au})$, and $(unfreeze,id, b+p_S, id_\mathcal{S})$, send $(check\_pass,id)$. 
						
						\item If any of the crtieria fails, call $\mathcal{L}$ with $(unfreeze, id,b+\frac{a}{2}, id_\mathcal{S})$ and $(unfreeze,id, p_S+p_{Au}+\frac{a}{2}, id_\mathcal{O})$. Send $(penalized,id,id_{Au})$.
						
					\end{itemize}
					\end{itemize}

\end{minipage}
}
\end{center}
\caption{Ideal Functionality $\mathcal{G}^{\mathcal{L}}_{\textsf{JC}}$ for the judge contract}\label{judgec} 
\end{figure*}
\begin{figure*}[!ht]
\ContinuedFloat
	\begin{center}
		\fbox{
			\begin{minipage}{1\textwidth}\
\textbf{(Server-Auditor: Closing Channel)}
 
				\begin{itemize}

					\item Receive $(complaint, id,Q,sig_Q,(\mu,\sigma),id_{\mathcal{S}})$ from $Au$ 
					\begin{itemize}
					\item Send $(get\_latest\_state,id)$ to $\mathcal{L}$, gets the value returned by $\mathcal{L}$, and stores the value in variable $seed$.
					\item Check $SigVerify(Q,sig_Q,pk_a)\stackrel{?}{=}1$. Generate $|Q|$ many queries from $seed$, and obtain $Q'$. If $Q!=Q'$, then go to step \emph{Penalize1}.
						\item Else, parse $Q$ to get $(i,\nu_i)$, where $i$ is the index of the file chunk. If $\mu\neq \phi$ and $\sigma\neq \phi$, check $e(\sigma,g)\stackrel{?}{=}e(\Pi_{(i,\nu_i)\in Q} H(i)^{\nu_i}u^{\mu},v)$. If this does not hold or if any of the response is empty, go to \emph{Wait for Response}. Else go to step \emph{Penalize1}.
					\item  \emph{Wait for Response}: If $\mathcal{S}$ sends $(\mu',\sigma'):e(\sigma',g)\stackrel{?}{=}e(\Pi_{(i,\nu_i)\in Q} H(i)^{\nu_i}u^{\mu'},v)$, then go to \emph{Penalize1}. If \emph{S} doesn't respond or the equation fails, go to step \emph{Penalize2}. 
					
					\item \emph{Penalize1}: Call $\mathcal{L}$ with $(unfreeze,id, \frac{a}{2}+b, id_{\mathcal{S}})$, and $(unfreeze,id, \frac{a}{2}+p_S+p_{Au}, id_\mathcal{O})$. Send $(penalized,id,id_{Au})$.

						\item \emph{Penalize2}: Call $\mathcal{L}$ with $(unfreeze,id, \frac{b}{2}+a, id_{Au})$ and $(unfreeze,id, \frac{b}{2}+p_S+p_{Au}, id_\mathcal{O})$. Send $(penalized,id,id_{\mathcal{S}})$.

		\end{itemize}
\item Receive $(complaint, id,Q,sig_Q,id_{Au})$ from $\mathcal{S}$ 
					\begin{itemize}
					\item Send $(get\_latest\_state,id)$ to $\mathcal{L}$, gets the value returned by $\mathcal{L}$, and stores the value in variable $seed$.
					\item Check $SigVerify(Q,sig_Q,pk_a)\stackrel{?}{=}1$. Generate $|Q|$ many queries from $seed$, and obtain $Q'$. If $Q!=Q'$ or signature check fails then call $\mathcal{L}$ with $(unfreeze, id,b+\frac{a}{2}, id_{\mathcal{S}})$ and $(unfreeze,id, p_S+p_{Au}+\frac{a}{2}, id_\mathcal{O})$. Send $(penalized,id,id_{Au})$.
						\item If comparison check holds true and signature is valid, call $\mathcal{L}$ with $(unfreeze,id, \frac{b}{2}+a, id_{Au})$ and $(unfreeze,id, p_S+p_{Au}+\frac{b}{2}, id_\mathcal{O})$. Send $(penalized,id,id_{\mathcal{S}})$.
						
		\end{itemize}
				\end{itemize}
\end{minipage}
}
\end{center}
\caption{Ideal Functionality $\mathcal{G}^{\mathcal{L}}_{\textsf{JC}}$ for the judge contract (Contd.)} 
\end{figure*}

The protocol $\Pi_{Cumulus}$ interacts with an instance of the global ledger $\mathcal{L}$ and the judge contract defined by ideal functionality $\mathcal{G}^{\mathcal{L}}_{\textsf{JC}}$. Any off-chain communication between any two honest parties occurs using ideal functionality for secure message transmission, $\mathcal{F}_{smt}$  \cite{canetti2000security}. The protocol is defined in Fig.\ref{ff1}. It calls the judge contract defined in Fig.\ref{judgec}. We mention the difference in the steps of the protocol for the privacy-preserving version in Fig.\ref{ff2} and the corresponding changes in the judge contract in Fig. \ref{judgecpriv}.

\begin{figure*}[!ht]
\begin{center}
\fbox{
\begin{minipage}{1.04\textwidth}\
The protocol consists of description of the behavior of honest owner $\mathcal{O}$, server $\mathcal{S}$ and auditor $Au$. It is assumed that the parties has access to global random oracle $\mathcal{H}$ and global ledger $\mathcal{L}$.\\
\textbf{Init Variables}: $u,g$ are generators of group $G$ of order $p$. \\

\textbf{(Initialize)}
\begin{itemize}
	\item Upon receiving $(store, id, F, p_S, p_{Au}, n, query\_count, id_{Au}, id_{\mathcal{S}})$, $\mathcal{O}$ selects the group $G$ based on $\lambda$, generators $u,g \in G$. $\mathcal{O}, \mathcal{S}$ and $Au$ calls Keygen function $\mathcal{F}_{ecdsa}$ and gets their secret key and public key pairs. The public key of $\mathcal{O},\mathcal{S}$ and $Au$ are $pk_o,pk_s,$ and $pk_a$.
	\item $\mathcal{O}$ sends $(init,id,pk_o, pk_{a},pk_s, u,v,g,\textrm{contract terms}, query\_count, id_{\mathcal{S}}, id_{Au})$ to $\mathcal{G}^{\mathcal{L}}_{\textsf{JC}}$. If it returns $(initialized, registered \ id_{\mathcal{O}}, registered \ id_{\mathcal{S}}, registered \ id_{Au})$, then continue, else abort.
	\end{itemize}
	\textbf{(File transfer from Owner to Server)}
\begin{itemize}

	\item $\mathcal{O}$ divides $F$ into $n$ blocks, $m_1,m_2,\ldots,m_n$, where each $m_i \in \mathbb{Z}_p, i \in [1,n]$.
	\item It generates authentication tags $\sigma_i=[\mathcal{H}(i)u^{m_i}]^{x}, 1\leq i \leq n$ and calculate $h_i=\mathcal{H}(m_i||\sigma_i), 1 \leq i \leq n$. Send $(m_i,\sigma_i)$ to $\mathcal{S}$.
	\item $\mathcal{S}$ checks $e(g,\sigma_i)\stackrel{?}{=}e(v,H(i)u^{m_i}), \forall 1 \leq i \leq n$. If the check holds for all the file chunks, it generates $h_i'=\mathcal{H}(m_i||\sigma_i)$ and signs by calling $t=Sign(\mathcal{O},[h_1',h_2',\ldots,h_n'])$ of $\mathcal{F}_{ecdsa}$, sends $([h_1',h_2',\ldots,h_n'],t)$ to $\mathcal{O}$.
	\item $\mathcal{O}$ calls $SigVerify([h_1',h_2',\ldots,h_n'],t,pk_s)$ of $\mathcal{F}_{ecdsa}$, checks if value returned is $1$, and $h_i'\stackrel{?}{=}h_i, i\in [1,n]$. If any check fails, abort. 
	\item $\mathcal{O}$ signs by calling $\sigma_{file}=Sign(\mathcal{O},([h_1',h_2',\ldots,h_n'],t))$ of $\mathcal{F}_{ecdsa}$, and sends $(verify,id,[h_1',h_2',\ldots,h_n'],t,\sigma_{file})$ to $\mathcal{G}^{\mathcal{L}}_{\textsf{JC}}$. If it responds with $(verified,id)$, then continue, else abort.
	\item $\mathcal{O}$ sends $(g,u,v,pk_o)$ to $Au$ and call $\mathcal{L}$ with $(freeze, id,p_S+p_{Au}, id_{\mathcal{O}})$. $Au$ sends an acknowledgement.
\end{itemize}    
\textbf{(Server-Auditor: Opening Channel)}
\begin{itemize}
				    \item $Au$ sends $( open\_channel, id, a, id_{\mathcal{S}})$ to $\mathcal{S}$ and waits for 1 round to receive acknowledgement from $\mathcal{S}$. 
					\item If $\mathcal{S}$ sends $( open\_channel, id, b, id_{Au})$ to $Au$, the latter sends $(open\_channel,id, a,id_{Au},b,id_{\mathcal{S}})$ to $\mathcal{G}^{\mathcal{L}}_{\textsf{JC}}$. If $\mathcal{S}$ doesn't respond in this round,  $\mathcal{O}$ calls $\mathcal{L}$ with $(unfreeze, id,p_S+p_{Au}, id_{\mathcal{O}})$ and abort.
					\item If $\mathcal{G}^{\mathcal{L}}_{\textsf{JC}}$ responds $(opened\_channel,id)$, then continue, else $\mathcal{O}$ calls $\mathcal{L}$ with $(unfreeze, id,p_S+p_{Au}, id_{\mathcal{O}})$ and abort.
					\item $\mathcal{S}$ sends $(get\_latest\_state,id)$ to $\mathcal{L}$, gets the value returned by $\mathcal{L}$, and stores the value in variable $seed$.
					\item $Au$ and $\mathcal{S}$, each initializes $Q_{all}=\mu_{all}=\sigma_{all}=\phi$ and $firstFlag=1$.
				\end{itemize}
	\textbf{(Server-Auditor: Query and Response)}
	
	\begin{itemize}
	    \item \emph{Query Phase}: $Au$ sends $(query,id,Q,sig_Q,id_{\mathcal{S}})$ to $\mathcal{S}$.
	    \item $\mathcal{S}$ adds $Q_{all}=Q_{all}\cup Q$, check if $Q$ can be generated from $seed$. If $Q$ cannot be generated from $seed$ or if $SigVerify(Q,sig_Q,pk_a)$ of $\mathcal{F}_{ecdsa}$ return 0, it sends $(complaint, id, Q,sig_Q,id_{Au})$ to $\mathcal{G}^{\mathcal{L}}_{\textsf{JC}}$ and terminates the protocol.
	    \item If the above criteria holds, $\mathcal{S}$ parses $Q$ and retrieves $\{(i,\nu_i)\}$, generates $\sigma = \prod_{(i,\nu_i)\in \mathcal{Q}} {\sigma_i}^{\nu_i}$,
      $\mu = \sum_{(i,\nu_i)\in \mathcal{Q}} \nu_i m_i$, creates $R=(\sigma,\mu)$, sends $(response, id, R,id_{\mathcal{S}})$  to $Au$. It adds $\mu_{all}=\mu_{all}+ \mu$ and $\sigma_{all}=\sigma_{all}\times \sigma$. 
      \item $Au$ parses $R$, gets $(\sigma,\mu)$, verifies whether $e(\sigma, g) \stackrel{?}{=} e(\prod_{(i,\nu_i)\in Q} H(i)^{\nu_i} u^{\mu}, v)$. If the check fails, it sends $( complaint,id, Q,sig_Q,(\mu,\sigma),id_{\mathcal{S}})$ to $\mathcal{G}^{\mathcal{L}}_{\textsf{JC}}$. Else it may either go back to \emph{Query Phase} and repeat the steps, or it may send $( close,id, Q_{all},sig_{Q_{all}},(\mu_{all},\sigma_{all}))$ to $\mathcal{G}^{\mathcal{L}}_{\textsf{JC}}$.

	\end{itemize}
    	\textbf{(Server-Auditor: Closing Channel)}
								\begin{itemize}
					    \item If the output is $(penalized,id,id_{\mathcal{S}})$ or $(penalized,id,id_{Au})$, it is sent to $id_{\mathcal{O}}$, $id_{\mathcal{S}}$ and $id_{Au}$, and the protocol terminates. 
					    \item If $Au$ receives $(check\_pass,id)$, then it sends it to $id_{\mathcal{O}}$. The latter now sends $(fetch\_file,id)$ to $id_{\mathcal{S}}$. The latter gets the file $F$ and terminates the protocol.
					\end{itemize}

\end{minipage}
}
\end{center}
\caption{The formal description of the protocol} 
\label{ff1}
\end{figure*}
\begin{figure*}[!ht]
\label{formal}
\begin{center}
\fbox{
\begin{minipage}{1.04\textwidth}\

\textbf{(Server-Auditor: Closing Channel)}
 
				\begin{itemize}
					\item Receive $(close,id, Q,sig_Q,(R,\mu,\sigma))$ from $Au$
					\begin{itemize}
						\item Parse $Q$ to get $(i,\nu_i)$, where $i$ is the index of the chunk.  Generate $W_i = (file\_id||i) \ \forall i \in Q$ \;,
      verify whether $R . e((\sigma)^\gamma, g) \stackrel{?}{=} e((\prod_{(i,\nu_i)\in \mathcal{Q}} H(W_i)^{\nu_i})^\gamma . u^{\mu}, v)$. If this holds true and $\frac{|Q|}{query\_count}\geq 1$, call $\mathcal{L}$ with $(unfreeze,id, a+p_{Au}, id_{Au})$, and $(unfreeze,id, b+p_S, id_\mathcal{S})$, send $(check\_pass,id)$. 
						
						\item If any of the crtieria fails, call $\mathcal{L}$ with $(unfreeze, id,b+\frac{a}{2}, id_\mathcal{S})$ and $(unfreeze,id, p_S+p_{Au}+\frac{a}{2}, id_\mathcal{O})$. Send $(penalized,id,id_{Au})$.
						
					\end{itemize}

					\item Receive $(complaint, id,Q,sig_Q,(R,\mu,\sigma),id_{\mathcal{S}})$ from $Au$ 
					\begin{itemize}
					\item Send $(get\_latest\_state,id)$ to $\mathcal{L}$, gets the value returned by $\mathcal{L}$, and stores the value in variable $seed$.
					\item Check $SigVerify(Q,sig_Q,pk_a)\stackrel{?}{=}1$. Generate $|Q|$ many queries from $seed$, and obtain $Q'$. If $Q!=Q'$, then go to step \emph{Penalize1}.
						\item Else, parse $Q$ to get $(i,\nu_i)$, where $i$ is the index of the file chunk. Generate $W_i = (file\_id||i) \ \forall i \in Q$ \;,
      verify whether $R . e((\sigma)^\gamma, g) \stackrel{?}{=} e((\prod_{(i,\nu_i)\in \mathcal{Q}} H(W_i)^{\nu_i})^\gamma . u^{\mu}, v)$. If this does not hold or if any of the response is empty, go to \emph{Wait for Response}. Else go to step \emph{Penalize1}.
					\item  \emph{Wait for Response}: If $\mathcal{S}$ sends $(R,\mu',\sigma')$:  generate $W_i = (file\_id||i) \ \forall i \in Q$ \;,
      verify whether $R . e((\sigma)^\gamma, g) \stackrel{?}{=} e((\prod_{(i,\nu_i)\in \mathcal{Q}} H(W_i)^{\nu_i})^\gamma . u^{\mu}, v)$\;. 
      If this holds true, then go to \emph{Penalize1}. If \emph{S} doesn't respond or the above equation fails, go to step \emph{Penalize2}. 
					
					\item \emph{Penalize1}: Call $\mathcal{L}$ with $(unfreeze,id, \frac{a}{2}+b, id_{\mathcal{S}})$, and $(unfreeze,id, \frac{a}{2}+p_S+p_{Au}, id_\mathcal{O})$. Send $(penalized,id,id_{Au})$.

						\item \emph{Penalize2}: Call $\mathcal{L}$ with $(unfreeze,id, \frac{b}{2}+a, id_{Au})$ and $(unfreeze,id, \frac{b}{2}+p_S+p_{Au}, id_\mathcal{O})$. Send $(penalized,id,id_{\mathcal{S}})$.

		\end{itemize}
					\end{itemize}

\end{minipage}
}
\end{center}
\caption{Ideal Functionality $\mathcal{G}^{\mathcal{L}}_{\textsf{JC}}$ for the judge contract (privacy preserving)} 
\label{judgecpriv}
\end{figure*}
\begin{figure*}[!ht]
\label{formal}
\begin{center}
\fbox{
\begin{minipage}{1.04\textwidth}\

	\textbf{(Server-Auditor: Query and Response)}
	
	\begin{itemize}
	    \item \emph{Query Phase}: $Au$ sends $(query,id,Q,sig_Q,id_{\mathcal{S}})$ to $\mathcal{S}$.
     \item $\mathcal{S}$ adds $Q_{all}=Q_{all}\cup Q$, check if $Q$ can be generated from $seed$. If $Q$ cannot be generated from $seed$ or if $SigVerify(Q,sig_Q,pk_a)=0$, it sends $(complaint, id, Q,sig_Q,id_{Au})$ to $\mathcal{G}^{\mathcal{L}}_{\textsf{JC}}$ and terminates the protocol.
	 
	    \item If the above criteria satisfies, $\mathcal{S}$ parses $Q$ and retrieves $\{(i,\nu_i)\}$, generates $\sigma = \prod_{(i,\nu_i)\in \mathcal{Q}} {\sigma_i}^{\nu_i}$, $R = e(u,v)^r$ \;, $\gamma = h(R)$ \;\\
      \eIf {$firstFlag==1$} {
      $\mu = r + \gamma \times \sum_{(i,\nu_i)\in \mathcal{Q}} \nu_i m_i$ \;\\
      $firstFlag=0$\;
      }
      {
      $\mu = \gamma \times \sum_{(i,\nu_i)\in \mathcal{Q}} \nu_i m_i$ \;
      }
      
    creates $\hat{R}=(R,\sigma,\mu)$, sends $(response, id, \hat{R},id_{\mathcal{S}})$  to $Au$. It adds $\mu_{all}=\mu_{all}+ \mu$ and $\sigma_{all}=\sigma_{all}\times \sigma$. 
      \item $Au$ parses $\hat{R}$, gets $(R,\sigma,\mu)$\;, 
       generate $W_i = (file\_id||i) \ \forall i \in Q$ \;,
      verify whether $R . e((\sigma)^\gamma, g) \stackrel{?}{=} e((\prod_{(i,\nu_i)\in \mathcal{Q}} H(W_i)^{\nu_i})^\gamma . u^{\mu}, v)$.\;If the check fails, it sends $( complaint,id, Q,sig_Q,(R,\mu,\sigma),id_{\mathcal{S}})$ to $\mathcal{G}^{\mathcal{L}}_{\textsf{JC}}$. Else it may either go back to \emph{Query Phase} and repeat the steps, or it may send $( close,id, Q_{all},sig_{Q_{all}},(R,\mu_{all},\sigma_{all}))$ to $\mathcal{G}^{\mathcal{L}}_{\textsf{JC}}$.

	\end{itemize}

\end{minipage}
}
\end{center}
\caption{The formal description of the protocol (Privacy-Preserving version). We just mention the difference of this protocol from the version mentioned in Fig. \ref{ff1}} 
\label{ff2}
\end{figure*}

\section{Security Analysis}
We show that any attack that can be performed on our protocol can also
be simulated on $\mathcal{F}^{\mathcal{L}}_{\textsf{Audit}}$, or in other words that our protocol is at least as secure as $\mathcal{F}^{\mathcal{L}}_{\textsf{Audit}}$. To prove this, we design a simulator \emph{Sim}, that acts like an ideal attacker for the ideal functionality. We show that no \texttt{PPT} environment can
distinguish between interacting with the real world and interacting with the ideal world. In the real world, the environment $\mathcal{Z}$ sends instructions to a real attacker $\mathcal{A}$ and interacts with our protocol. In the ideal world, $\mathcal{Z}$ sends attack instructions to \emph{Sim} and interacts with $\mathcal{F}^{\mathcal{L}}_{\textsf{Audit}}$.

\emph{UC Definition of security}. Consider the protocol \emph{Cumulus}, denoted as $\Pi_C$ with access to the judge contract functionality $\mathcal{G}^{\mathcal{L}}_{JC}$, the global random oracle $\mathcal{H}$, ECDSA-signature functionality $\mathcal{F}_{ecdsa}$, and the global ledger functionality $\mathcal{L}$. Let $REAL_{\Pi_C,\,mathcal{A},\mathcal{Z}}^{\mathcal{F}_{ecdsa},\mathcal{G}^{\mathcal{L}}_{JC},\mathcal{H}}(\lambda,x)$ be the ensemble of the outputs of the environment $\mathcal{Z}$ when interacting with the attacker $\mathcal{A}$ and users running protocol $\Pi_C$ on input $1^{\lambda}$, where $\lambda$ is the security parameter, and auxiliary input $x \in \{0,1\}^*$, in the hybrid world having access to the ideal functionalities. The parties interact with each other via $\mathcal{F}^{\mathcal{L}}_{\textsf{Audit}}$ in the ideal world in the presence of the ideal attacker $Sim$. $IDEAL_{Sim,\mathcal{Z}}^{\mathcal{F}^{\mathcal{L}}_{\textsf{Audit}},\mathcal{H},\mathcal{L}}(\lambda,x)$ be the ensemble of the outputs of the environment $\mathcal{Z}$ when interacting with the ideal world.

\begin{theorem}
\textbf{Global UC Security}. Given that $\lambda$ is the security parameter and ECDSA signatures are strongly existentially unforgeable, a protocol $\Pi_C$ is said to GUC-realizes an ideal functionality $\mathcal{F}^{\mathcal{L}}_{\textsf{Audit}}$ in the $(\mathcal{F}_{ecdsa},\mathcal{G}^{\mathcal{L}}_{JC},\mathcal{H})$-hybrid world if for all computationally bounded adversary $\mathcal{A}$ attacking $\Pi_C$, there exist a probabilistic polynomial time (PPT) simulator $Sim$ such that for all PPT environment $\mathcal{Z}$ and for all $x \in \{0,1\}^*$, $IDEAL_{Sim,\mathcal{Z}}^{\mathcal{F}^{\mathcal{L}}_{\textsf{Audit}},\mathcal{H},\mathcal{L}}(\lambda,x)$, and $REAL_{\Pi_C,\,mathcal{A},\mathcal{Z}}^{\mathcal{F}_{ecdsa},\mathcal{G}^{\mathcal{L}}_{JC},\mathcal{H}}(\lambda,x)$ are computationally indistinguishable.
\end{theorem}
\textit{Proof}: To prove that our proposed protocol is \emph{GUC-secure}, we need to show that any PPT environment $\mathcal{Z}$ can distinguish the execution of the protocol in hybrid world from the ideal world with negligible probability. We construct a simulator $Sim$ that outputs all messages such that it looks like the hybrid world execution to $\mathcal{Z}$. The first version of \emph{Cumulus}, defined in Fig.\ref{ff1}, is more efficient but has a weaker privacy guarantee, the second version defined in Fig.\ref{ff2} is more privacy-preserving but lacks efficiency. $Sim$ orchestrates the response depending on which version of the protocol it is trying to simulate in the ideal world.

The transcript of the protocol generated in the hybrid world must not be distinguishable even in presence of corrupted parties. We consider total of eight cases, the protocol execution with all honest parties, execution with a malicious owner $\mathcal{O}^*$, execution with a dishonest server $\mathcal{S}^*$, execution with dishonest auditor $Au$, three cases for execution with either of two parties in set $\{\mathcal{O},\mathcal{S},Au\}$ being dishonest, and the case where all parties are corrupt. 

We consider that $Sim$ internally simulates the execution of $\mathcal{G}^{\mathcal{L}}_{\textsf{JC}}$, suppressing any calls made to ledger $\mathcal{L}$ and has access to the global random oracle $\mathcal{H}$ and $\mathcal{F}_{ecdsa}$ for signing messages on behalf of each party. We assume that $u,g$ are public parameters.
\begin{itemize}
    \item \emph{Simulation without corruptions}: The simulator $Sim$, defined in Fig.\ref{sim:honest}, is just required to match the transcript of all messages of the execution of \emph{Cumulus} in the hybrid world. Since all three parties are honest, none of the messages exchanged between these parties is leaked to $\mathcal{Z}$. Since $Sim$ has no access to the file $F$, it generates a random bit string having length same as the size of the file, and performs all the operations on the dummy file. $\mathcal{Z}$ who has no access to the file, cannot distinguish between the output of $Sim$ and output by an auditor upon channel closure.
  
    \item \emph{Simulation when $\mathcal{O}$ is dishonest}: The simulator is defined in Fig.\ref{sim:o-corrupted}. The dishonest owner will be able to generate $h_i', 1\leq i \leq n$ such that it matches with the hash of the chunks had it possessed the correct file with probability $\frac{1}{p^n}$. This is the only bad case where $Sim$ aborts but the protocol execution in hybrid world will continue. But since the probability is negligible, the two executions can be distinguished with negligible probability.

        \item \emph{Simulation when $\mathcal{S}$ is dishonest}: We define the simulator for a dishonest server in Fig. \ref{sim:s-corrupted}. In the first step, even if the dishonest sever manages to generate tag $\sigma_i$ without actually having the message $m_i$, it will never pass the check $h_i = \mathcal{H}(m^*||\sigma_i)$ if $m^*\neq m_i$. Since we consider ECDSA signature to be existentially unforgeable, the malicious sender cannot forge a signature without posessing the correct secret key.
        
        In the auditing phase, given a query $Q$, the sender may try to guess the response $(\mu',\sigma')$. Probability of guessing the tuple correctly is $\frac{1}{p^2}$, where $Pr[\mu=\mu']=\frac{1}{p}$ and $Pr[\sigma=\sigma']=\frac{1}{p}$. This is a bad event and $Sim$ aborts if this event occurs. Since $\frac{1}{p^2}$ is a negligible quantity, probability of $Sim$ aborting is negligible. Since we consider the signature scheme is secure, the probability of this bad event occurring is negligible.

            \item \emph{Simulation when $Au$ is dishonest}: We define the simulator for a dishonest server in Fig. \ref{sim:au-corrupted}. The bad case can arise in two situation: (i) $Au$ raises a false complaint that $\mathcal{S}$ has send a wrong query. However, in that case, $Au$ needs to forge $\mathcal{S}'s$ signature as well and this is possible with negligible probability. (ii) The next bad case arises when $Au$ submits a channel closure request but without the correct response. This will be detected by the ideal judge contract and probability of the bad event is 0. Also, it is possible that the malicious auditor has not interacted with server and tries to generate response for a given query set. Since we show that probability of guessing a such a tuple is $\frac{1}{p^2}$ in the previous instance where $\mathcal{S}$ was dishonest, so the bad event is possible with negligible probability.   

               \item \emph{Simulation when $\mathcal{O},\mathcal{S}$ are dishonest}:  We define the simulator for a dishonest server in Fig. \ref{sim:os-corrupted}. The only point where both the parties can collude and try to steal coins from an honest $Au$ is when they falsely raise a complaint about a wrong query. However, ECDSA signature is existentially unforgeable, the attack is possible only with negligible probability.
                 \item \emph{Simulation when $\mathcal{O},Au$ are dishonest}: In no way can the two dishonest parties cheat an honest sender who posseses the correct response to each query. We define the simulator in Fig. \ref{sim:oau-corrupted}.  
                \item \emph{Simulation when $\mathcal{S},Au$ are dishonest}: We define the simulator for a dishonest server in Fig. \ref{sim:aus-corrupted}. It is possible that the malicious auditor-server pair tries to generate response for a given query set. Probability of guessing the tuple correctly is $\frac{1}{p^2}$, where $Pr[\mu=\mu']=\frac{1}{p}$ and $Pr[\sigma=\sigma']=\frac{1}{p}$. This is a bad event and $Sim$ aborts if this event occurs. Since $\frac{1}{p^2}$ is a negligible quantity, probability of $Sim$ aborting is negligible.

\item \emph{Simulation when all parties are dishonest}: With all three parties behaving dishonestly, the action set is exponential in size and cannot be captured here. This case does not guarantee any fairness and might not even
terminate. We therefore skip discussions for this bad case. 

\end{itemize}

\begin{figure*}[!ht]
\label{formal}
\begin{center}
\fbox{
\begin{minipage}{1.04\textwidth}\

\begin{itemize}
\item $Sim$ receives $(|F|,n,p_S,p_{Au},seed,id_{\mathcal{S}}, id_{\mathcal{O}},id_{Au})$ from $\mathcal{F}^{\mathcal{L}}_{\textsf{Audit}}$. 
\begin{itemize}
    \item It samples $(sk_o,pk_o), (sk_s,pk_s)$ and $(sk_a,pk_a)$, and simulates the execution of protocol by sending $(init,id,pk_o, pk_{a},pk_s, u,v,g,\textrm{contract terms}, query\_count, id_{\mathcal{S}}, id_{Au})$ to $\mathcal{G}^{\mathcal{L}}_{\textsf{JC}}$. 

\item It generates a file $F'=1^{|F|}$ and generates $m_1',m_2',\ldots,m_n'$ by breaking $F'$ into $n$ chunks of equal size. 
\item $Sim$ samples a secret key $x'$ and public key $v'=g^{x'}$ and creates $\sigma_i=[\mathcal{H}(i)u^{m_i'}]^{x'}, h_i=\mathcal{H}(m_i'||\sigma_i), i \in [1,n]$.
\item $Sim$ simulates these steps on behalf of $\mathcal{O}$ and $\mathcal{S}$. It signs $t=Sign([h_1,h_2,\ldots,h_n],sk_s)$, and again countersigns $\sigma_{file}=Sign(([h_1,h_2,\ldots,h_n],t),sk_o)$,  and simulates the execution of protocol by sending $(verify,id,[h_1,h_2,\ldots,h_n],t,\sigma_{file})$ to $\mathcal{G}^{\mathcal{L}}_{\textsf{JC}}$.
\end{itemize}
\item $Sim$ receives $( open\_channel, id, b, id_{Au})$.
\begin{itemize}
\item It simulates the execution of protocol by sending $(open\_channel,id, a,id_{Au},b,id_{\mathcal{S}})$ to $\mathcal{G}^{\mathcal{L}}_{\textsf{JC}}$.
\end{itemize}
\item $Sim$ receives $(query,id,Q)$:
\begin{itemize}
    \item If it observes $\mathcal{L}$ has received $(unfreeze, id,b+\frac{a}{2}, id_\mathcal{S})$ and $(unfreeze,id, p_S+p_{Au}+\frac{a}{2}, id_\mathcal{O})$, then it sends $( complaint,id, Q,sign_{sk_a}(Q),id_{Au})$ to $\mathcal{G}^{\mathcal{L}}_{\textsf{JC}}$.
    \item If it observes $\mathcal{L}$ has received $(unfreeze, id,a+\frac{b}{2}, id_{Au})$ and $(unfreeze,id, p_S+p_{Au}+\frac{b}{2}, id_\mathcal{O})$, then it sets $\mu=\Sigma_{(i,\nu_i)\in Q} \nu_i m_i'$, and $\sigma=\phi$, and sends $( complaint,id, Q,sign_{sk_a}(Q),(\mu,\sigma),id_{\mathcal{S}})$ to $\mathcal{G}^{\mathcal{L}}_{\textsf{JC}}$.
\end{itemize}
\item $Sim$ receives $( close,id, Q,\mu)$:
\begin{itemize}
    \item If it observes that $\frac{|Q|}{query\_count}\geq 1$, and $\mathcal{L}$ has received $(unfreeze, id,b+\frac{a}{2}, id_\mathcal{S})$ and $(unfreeze,id, p_S+p_{Au}+\frac{a}{2}, id_\mathcal{O})$, then it sets $\mu=\Sigma_{(i,\nu_i)\in Q} \nu_i m_i'$, and $\sigma=\phi$ and goes directly to step \emph{Call}. 

    \item Else, it parses $Q$ and gets $\{(i,\nu_i)\}$, generates $\mu=\Sigma_{(i,\nu_i)\in Q} \nu_i m_i'$ and $\sigma=\Pi_{(i,\nu_i) \in Q} \sigma_i^{\nu_i}$.
    \item \emph{Call}: $Sim$ signs the query $Q$ using $sk_a$, simulates the execution of protocol by sending $( close,id, Q,sign_{sk_a}(Q),(\mu,\sigma))$ to $\mathcal{G}^{\mathcal{L}}_{\textsf{JC}}$.
\end{itemize}
\end{itemize}

\end{minipage}
}
\end{center}
\caption{Simulator $Sim^{honest}$ for without corruption} 
\label{sim:honest}
\end{figure*}

\begin{figure*}[!ht]
\label{formal}
\begin{center}
\fbox{
\begin{minipage}{1.04\textwidth}\	
				\begin{itemize}
					\item Upon receiving $(init,id,p_S, p_{Au},pk_o, pk_{a},pk_s, u,v,g,\textrm{contract terms}, query\_count, id_{\mathcal{S}}, id_{Au})$ from $\mathcal{O}^*$, $Sim$ simulates execution by sending it to $\mathcal{G}^{\mathcal{L}}_{\textsf{JC}}$. 
     \item $\mathcal{O}^*$ sends $\{(m_i,\sigma_i): 1\leq i \leq n\}$, $Sim$ checks if $e(g,\sigma_i)=e(v,H(i)u^{m_i})$. If it holds true for all file chunks, it creates file $F^*=\{m_i:1\leq i \leq n \}$
     \item $\mathcal{O}^*$ sends $(verify,id,[h_1',h_2',\ldots,h_n'],t,\sigma_{file})$ to $Sim$. The latter checks if $h_i'=\mathcal{H}(m_i||\sigma_i), 1\leq i \leq n$. if there is a mismatch abort.
     \item $Sim$ simulates execution by sending the message to $\mathcal{G}^{\mathcal{L}}_{\textsf{JC}}$. If $\texttt{SigVerify($m$,$t$,$pk_s$)=1}$ and
 $\texttt{SigVerify($(m,t)$,$s,pk_o$)=1}$, send $(store, id, F^*, p_S, p_{Au}, query\_count, n,  id_{Au}, id_{\mathcal{S}})$ to $\mathcal{F}^{\mathcal{L}}_{\textsf{Audit}}$.
			\end{itemize}
\end{minipage}
}
\end{center}
\caption{Simulator $Sim^{\mathcal{O},corrupted}$ } 
\label{sim:o-corrupted}
\end{figure*}
\begin{figure*}[!ht]
\label{formal}
\begin{center}
\fbox{
\begin{minipage}{1.04\textwidth}\
\begin{itemize}
    \item $\mathcal{F}^{\mathcal{L}}_{\textsf{Audit}}$ sends file $F$ to $Sim$. The latter parses $F$ to $m_1,m_2,\ldots,m_n$.
\item $\mathcal{S}^*$ sends $\{\sigma_i: 1 \leq i \leq n\}$, $([h_1',h_2',\ldots,h_n'],t)$ to $Sim$. The latter checks if $e(g,\sigma_i)\stackrel{?}{=}e(v,H(i)u^{m_i})$, and $h_i=\mathcal{H}(m_i||\sigma_i), 1\leq i \leq n$, and $SigVerify([h_1',h_2',\ldots,h_n'],t,pk_s)\stackrel{?}{=}1$, if any of the check fails, send $(abort,id)$ to $\mathcal{F}^{\mathcal{L}}_{\textsf{Audit}}$.
\item Upon receiving $(open\_channel, id, a, id_{\mathcal{S}})$, $Sim$ forwards the message to $S^*$. If the latter responds with $(open\_channel, id, b, id_{Au})$, it initiates opening of channel. If no message is received, it sends $(abort,id)$ to $\mathcal{F}^{\mathcal{L}}_{\textsf{Audit}}$.
\item $\mathcal{S}^*$ sends $(\sigma',\mu')$ to $Sim$, the latter checks that for corresponding $\mathcal{Q}$, $\sigma'=\Pi_{(i,\nu_i)\in \mathcal{Q}}\sigma_i^{\nu_i}$, $\mu'=\sum_{(i,\nu_i)\in \mathcal{Q}}\nu_i m_i$.
\item If $S^*$ sends $(complaint,id,Q,sig_Q,id_{Au})$ to $Sim$, the latter checks if $Q$ is valid query generated from $seed$ and if $SigVerify(Q,sig_Q,pk_a)=1$. If any of the check fails, it sends $(abort,id)$ to $\mathcal{F}^{\mathcal{L}}_{\textsf{Audit}}$. 
\end{itemize}


\end{minipage}
}
\end{center}
\caption{Simulator $Sim^{\mathcal{S},corrupted}$ } 
\label{sim:s-corrupted}
\end{figure*}
\begin{figure*}[!ht]
\label{formal}
\begin{center}
\fbox{
\begin{minipage}{1.04\textwidth}\
\begin{itemize}
    \item If $Au^*$ sends $(complaint,id,Q,sig_Q,(\mu',\sigma'),id_{Au})$ to $Sim$, the latter checks if $Q$ is valid query generated from $seed$ and if $SigVerify(Q,sig_Q,pk_a)=1$. If any of the check fails, it sends $(abort,id)$ to $\mathcal{F}^{\mathcal{L}}_{\textsf{Audit}}$. If the check is valid, then $Sim$ checks if $\mathcal{L}$ has taken any action or remains inactive. If the latter happens then $Sim$ sends $(abort,id)$. If funds are unfrozen penalizing server $S$, then it does nothing. 
    \item If $Au^*$ sends $(close,id,Q,sig_Q,(\mu',\sigma'),id_{Au})$ to $Sim$, parse $Q$ to get $(i,\nu_i)$, where $i$ is the index of the chunk. Check $e(\sigma,g)\stackrel{?}{=}e(\Pi_{(i,\nu_i)\in Q} H(i)^{\nu_i}u^{\mu},v)$. If this is not true send $(abort,id)$ to $\mathcal{F}^{\mathcal{L}}_{\textsf{Audit}}$.
\end{itemize}

\end{minipage}
}
\end{center}
\caption{Simulator $Sim^{Au,corrupted}$} 
\label{sim:au-corrupted}
\end{figure*}

\begin{figure*}[!ht]
\label{formal}
\begin{center}
\fbox{
\begin{minipage}{1.04\textwidth}\
$\mathcal{O}^*$ and $\mathcal{S}^*$ collude and try to cheat $Au$. It is possible either by supplying a wrong $seed$ or raising a dispute on the query set $Q$.
\begin{itemize}
    \item Since $seed$ is present in the ideal judge contract, this attack can be ruled out.
    \item The latter case is possible, if they can forge $Au$'s signature. This is possible with negligible probability.
\end{itemize}

\end{minipage}
}
\end{center}
\caption{Simulator $Sim^{(\mathcal{O},\mathcal{S}),corrupted}$} \label{sim:os-corrupted}
\end{figure*}

\begin{figure*}[!ht]
\label{formal}
\begin{center}
\fbox{
\begin{minipage}{1.04\textwidth}\
$\mathcal{O}^*$ and ${Au}^*$ collude and try to cheat $\mathcal{S}$. 

\begin{itemize}
    \item Same as the case when $Au^{*}$ was corrupted. If ${Au}^*$ tries to suppress or send a wrong response, then judge contract will wait for the correct response from $\mathcal{S}$. If an honest $\mathcal{S}$ responds, then it will be paid accordingly.
\end{itemize}

\end{minipage}
}
\end{center}
\caption{Simulator $Sim^{(\mathcal{O},Au),corrupted}$ } 
\label{sim:oau-corrupted}
\end{figure*}
\begin{figure*}[!ht]
\label{formal}
\begin{center}
\fbox{
\begin{minipage}{1.04\textwidth}\
If $Au^*$ sends $(close,id,Q,sig_Q,(\mu',\sigma'),id_{Au})$ to $Sim$, parse $Q$ to get $(i,\nu_i)$, where $i$ is the index of the chunk. Check $e(\sigma,g)\stackrel{?}{=}e(\Pi_{(i,\nu_i)\in Q} H(i)^{\nu_i}u^{\mu},v)$. If this is not true send $(abort,id)$ to $\mathcal{F}^{\mathcal{L}}_{\textsf{Audit}}$.

\end{minipage}
}
\end{center}
\caption{Simulator $Sim^{(Au,\mathcal{S}),corrupted}$ } 
\label{sim:aus-corrupted}
\end{figure*}


\section{Implementation and Performance Analysis}
In this section, we analyze a realistic cloud setting of blockchain enabled data audit scheme that we have implemented.

\subsection{Implementation Setup}
We implement AuB and evaluate a prototype using \texttt{Ethereum} as a blockchain platform. Our entire code is approximately $1500$ lines, consisting of Ethereum smart contracts written in Solidity language, Go-Ethereum modifications written in Golang and other experimentation glue code written in Python and Bash.~\footnote{\href{https://bitbucket.org/prabal_banerjee/audit/src/master/}{https://bit.ly/2L6W55n} and \href{https://github.com/prabal-banerjee/go-ethereum.git}{https://bit.ly/2J0z1Ct}}\hfill

We needed to perform Bilinear Pairing checks for symmetric pairing inside Ethereum smart contract, to verify audit responses. In this regard, the original Ethereum platform does not support pairing based operations on symmetric groups natively. Post Byzantium, it had introduced pairing operations on a fixed asymmetric group, in order to support Zero-Knowledge proof verification. It was impractical to port some pairing-based cryptography library into Solidity and hence we modified the Ethereum code to include a new pre-compiled contract which supported pairing-based operations. To be specific, the new pre-compiled contract verified the audit equation given in Eq.\ref{verEqSW}.

We have used the most popular Ethereum implementation, Go Ethereum also known as \texttt{geth}, which is written in Golang. For the mathematical operations, we needed a library which supports arithmetic in $\mathbb{Z}_p$, elliptic curve groups with operations and bilinear pairing computation. Hence, we used the Golang wrapper \cite{pbcGolang} of the popular \texttt{PBC Library}\cite{Lynn}. We included the PBC library inside the geth code and introduced the pre-compiled contract. We used Type A pairings which are very fast, but elements take a lot of space to represent. Because of the modified Ethereum code, we used a private network for our experimentation. 

For the Shacham-Waters audit code implementation, we used their extended definition with file sectors. As discussed in algorithm\ref{transfer}, we split file $F$ into $n$ blocks $m_1, m_2,..., m_n \in \mathbb{Z}_p$. For each block $m_i$, tag $\sigma_i \in G$ is calculated, where $G$ is group whose support is $\mathbb{Z}_p$. Calculating tags for the main file $F$ causes a significant overhead if we generate tags as above. So, we used the concept of sectors as introduced by Shacham-Waters. Let $s$ be a parameter and each block consist of $s$ sectors, where $|s| \in \mathbb{Z}_p$. As there is only one tag for one block (contains $s$ sectors), tag generation overhead is reduced by $\approx1/s$ if $n$ is large enough.

\subsection{Evaluation}
We deployed our implementation on a private Ethereum network consisting of two nodes. We used a single machine with $4$-core Intel Xeon $E3-1200$ and $8$GB of RAM running Linux (Manjaro 64-Bit XFCE). The storage server, owner and auditor codes were running alongside the Ethereum nodes. The elliptic curve utilized in our experiment is a supersingular curve, with a base field size $512$ bits and the embedding degree $2$. We use different file sizes, starting from $1$KB to $100$MB. 

Sector size $|s|$, is $19$ Bytes in our construction which is dependent upon parameters we used for the construction of elliptic curves. We use 1000 sectors per block in our construction which we noticed is optimized value for current setup.

The main objective of our prototype implementation was to observe the overhead in introducing the distributed computing platform. In particular, we wanted to calculate latency from each of the parties perspective, i.e., how much additional time does the owner spend in uploading and downloading files, the auditor spends in challenge-response and the server spends on file and audit management. To observe this, we perform same experiment with and without the blockchain related calls and look into the latencies in each case. 

\begin{figure}[H]
	    \centering
		\captionsetup{justification=centering,font=scriptsize}
	    \includegraphics[width=0.7\textwidth]{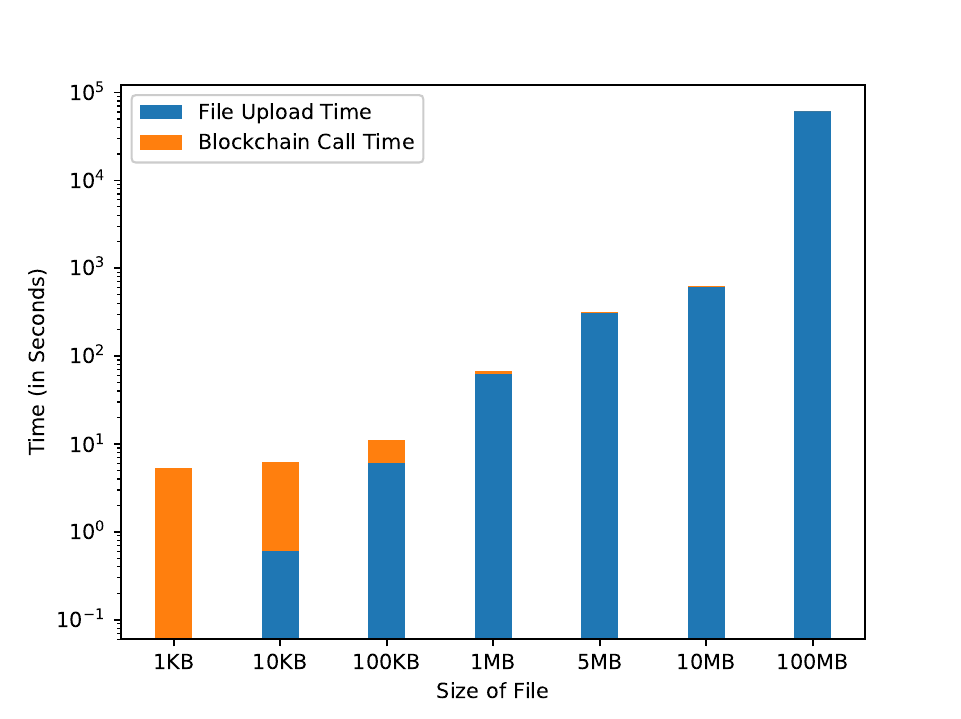}
		\caption{File upload time for different $|F|$}
		\label{fig:upload}
	\end{figure}

Firstly, we look into the latency faced by the owner during upload of file. In Fig. \ref{fig:upload}, we note that although for small files the blockchain latency remains considerable, with increasing file sizes, the commit time becomes negligible compared to the file upload time. As practical storage servers store files in order of Gigabytes, the overhead for the owner is negligible. A thing to note is that the same applies for the server as well because the owner latency includes the server signing during file uploads. 

\begin{figure}[H]
		\centering
		\captionsetup{justification=centering,font=scriptsize}
		\includegraphics[width=0.7\textwidth]{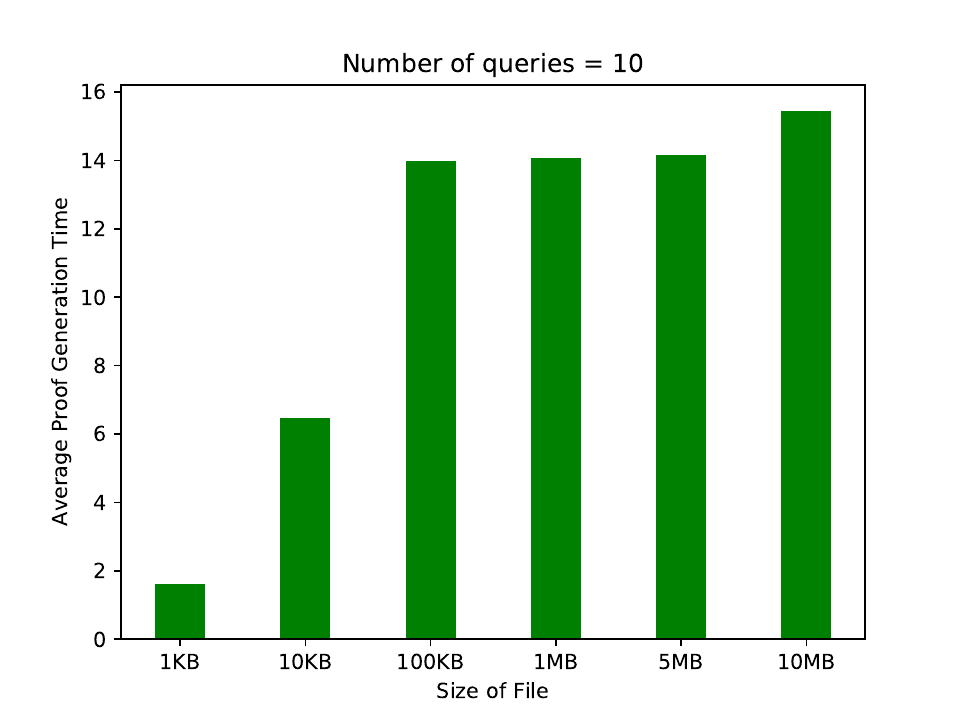}
		\caption{Proof generation time for different $|F|$}
		\label{fig:proofgen}
		\end{figure}

The protocol does not demand any additional ledger interactions during proof generation and hence we observed no significant overhead for the servers, as shown in Fig. \ref{fig:proofgen}.

\begin{figure}[H]
						\centering
						\captionsetup{justification=centering,font=scriptsize}
						\includegraphics[width=0.7\textwidth]{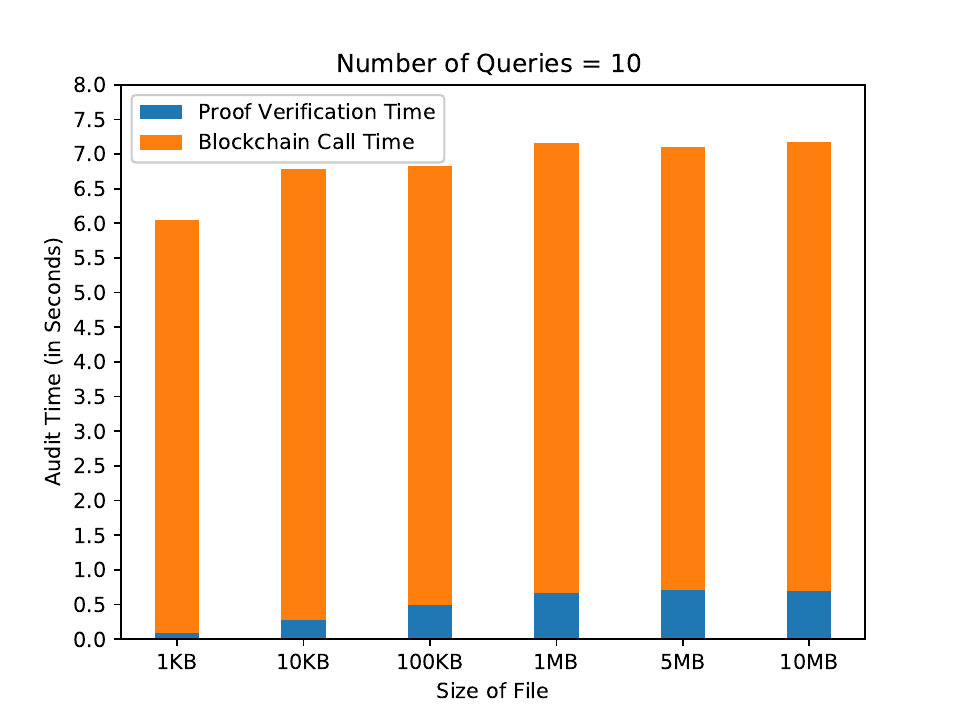}
						\caption{Proof verification time for different $|F|$} 
						\label{fig:proofver}
					\end{figure}

An auditor performs audit over a long period of time. For example, an auditor may send one query to the server every hour. It may have to send the aggregated response to the smart contract only at the end of the day. Hence, in Fig. \ref{fig:proofver} although we observe a dominant overhead of blockchain interaction compared to response verification over 10 queries, we note that the time axis in not an honest representation of practice where the audit will be performed over a considerably long period of time as compared to the commit time. 

\begin{figure}[H]
						\centering
						\captionsetup{justification=centering,font=scriptsize}
						\includegraphics[width=0.7\textwidth]{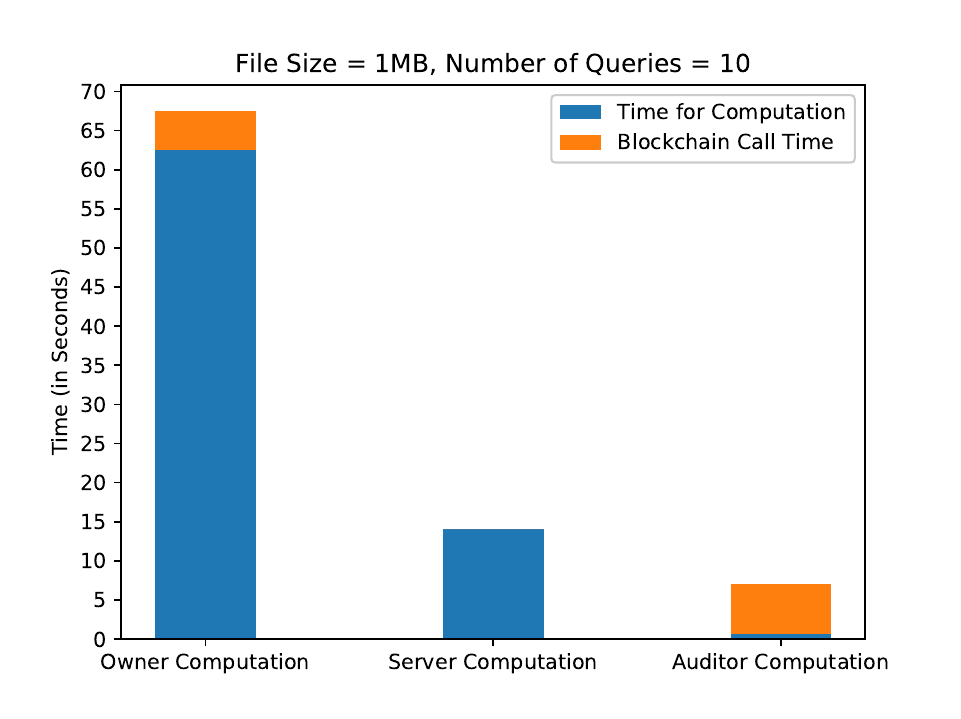} 
						\caption{Computational time for all parties} 
						\label{fig:verification}
					\end{figure}

Overall, in Fig. \ref{fig:verification}, we see that for the owner and server, our protocol adds minimal overhead. For the auditor, if it is compared against the span of the entire audit process, the additional latency remains negligible, given that the auditor performs the audit over a sufficiently long duration of time.

Table \ref{table:1} shows the metrics calculated with $5$ different query sizes keeping the file size constant at 1 MB. The gas cost in USD is calculated   
at average gas price of 3 gwei and an exchange rate of 1 ETH = 153 USD. The empty block size in our private network is 540 bytes. This shows that if a single audit takes 1400 bytes, 1 MB data on the blockchain would accommodate roughly 750 audits. We note here that the block size increase is not linear to the number of queries as only the aggregated response is submitted to the blockchain. Each channel session can communicate a large number of audits hence in practice, thousands of such audits can be done with an overhead of few kilobytes.  

\textit{In comparison to previous work like \cite{zkcp}, not only is our proof generation time lower, our proof size is also smaller as it is aggregated over hundreds of proofs over time. This is because of the off-chain nature of our solution.}

\begin{table*}[!htbp]
\hspace*{-1cm}
\centering
\begin{tabular}{llllll}
\#Queries & Gas Cost & Gas Cost (\$) & Block Overhead & Auditing Time & BC Time \\
 \hline
 1 & 110087 & $0.0512$ & 1394 B & 72.808527ms & 6.220183447s \\
 3 & 115220 & $0.05358$ & 1458 B & 65.074692ms & 6.215553521s \\
 5 & 120486 & $0.05603$ & 1554 B & 69.638309ms & 7.229788269s \\
 7 & 125619 & $0.05842$ & 1618 B & 67.183777ms & 5.818474706s \\
 10 & 133451 & $0.06206$ & 1746 B & 69.449573ms & 6.219303512s
\end{tabular}
\caption{Cost of audit for file size of $1$MB.}
\label{table:1}
\end{table*}



\section{Discussion}
We wanted to use blockchain as the source of randomness for generating query set. As given in \cite{bitcoinrandomness}, for small amounts of randomness, if the stakes are low enough, the blockchain can be used as a source of randomness. We believe that for our audit purposes, the incentive for parties to collude with miners is low enough. Any other public source of randomness could have been used. External sources of randomness have a separate trust assumption and then we would have needed to consider all the collusion cases with the random source. Our only requirement is that the peers of the blockchain network need to have access to the same source and must access the same random value in order to receive consensus. We referred a single block hash for each contract instance, hence the query set for a channel can be derived at once, after the opening of the channel. 

File upload time is very much dependent on number of sectors per block as well as size of sector. Sector size $|s|$, is determined by the parameters and choice of algorithm used for elliptic curve generation. 

We have implemented our pairing check as a new pre-compiled contract. Hence, the gas required by the contract has been estimated by us. In a practical situation, either such a symmetric bilinear pairing support comes baked into Ethereum, in which case the community decides upon the gas cost, or, a private network is setup among interested parties where they themselves decide upon the gas requirement. The asymmetric pairing check pre-compiled contract takes $80000*k + 100000$ as the gas ($k$ is the number of points on the curve). Upon using similar calculation, our audit check transaction took $888387$ gas. We have not used this in our performance metric as we think this will depend upon the platform.
 
The channel closing codes written in our contract is far from ideal. It does not take into account all possible corner cases, but arbitrary complicated code could have been implemented based on the requirements. We have just showed a sample code for the prototype. Also, we note here that in a classical audit scenario, both auditor and server needs to be online during audit phase. For our proposed state channel to work, this is the exact requirement and hence it imposes no additional restrictions. 

\section{Conclusion}
In this paper we introduced a blockchain based privacy preserving audit protocol which is resilient even when any two out of the data owner, storage server and auditor is malicious. We used state channels to minimize blockchain commits thereby improving efficiency. Through smart contracts, we enforced the incentive mechanism in the system. We also build a prototype on modified Ethereum and show that the protocol incurs minimal overhead compared to existing PoR scheme. 

In terms of future work, we wish to explore possibilities to enhance efficiency of the protocol by using other elliptic curves. We also aim to adopt an audit protocol without bilinear pairing operations so that it can be readily deployed on blockchain platforms like Ethereum, without modifying the codebase. This would enable us to test on networks beyond a private network, like testnets and main network. 

\section*{Acknowledgment}
This work is partially supported by Cisco University Research Program Fund, CyberGrants ID: \#698039 and Silicon Valley Community Foundation. The authors would like to thank Chris Shenefiel and Samir Saklikar for their comments and suggestions. The work is also partially supported by the European Research Council (ERC) under the European Union’s Horizon 2020 research (grant agreement 771527- BROWSEC), by the Austrian Science Fund (FWF) through the projects PROFET (grant agreement P31621), and the project W1255- N23, by the Austrian Research Promotion Agency (FFG) through the COMET K1 SBA and COMET K1 ABC, by the Vienna Business Agency through the project Vienna Cybersecurity and Privacy Research Center (VISP), by the Austrian Federal Ministry for Digital and Economic Affairs, the National Foun- dation for Research, Technology and Development and the Christian Doppler Research Association through the Christian Doppler Laboratory Blockchain Technologies for the Internet of Things (CDL-BOT).


  \bibliographystyle{ACM-Reference-Format}
  \bibliography{ref}

\end{document}